\documentclass[twocolumn,preprintnumbers]{revtex4}
\usepackage{amsmath}
\usepackage{dcolumn}
\usepackage{bm}
\usepackage{epsfig}
\usepackage{epstopdf}
\usepackage{graphicx}
\usepackage{amsfonts}
\usepackage{amssymb}
\usepackage{mathrsfs}
\usepackage{color}
\usepackage{multirow}
\usepackage{appendix}
\usepackage{afterpage}
\newcommand\myemptypage{
	\null
	\thispagestyle{empty}
	\addtocounter{page}{-1}
	\newpage
}

\begin{document}
\title{Single-atom verification of the noise-resilient and fast characteristics of universal nonadiabatic noncyclic geometric quantum gates}
\author{J. W. Zhang$^{1,3}$}
\author{L.-L. Yan$^{2}$}
\author{J. C. Li$^{1,3}$}
\author{G. Y. Ding$^{1,3}$}
\author{J. T. Bu$^{1,3}$}
\author{L. Chen$^{1}$}
\author{S.-L. Su$^{2}$}
\email{slsu@zzu.edu.cn}
\author{F. Zhou$^{1}$}
\email{zhoufei@wipm.ac.cn}
\author{M. Feng$^{1,2,3,4}$}
\email{mangfeng@wipm.ac.cn}
\affiliation{$^{1}$ State Key Laboratory of Magnetic Resonance and Atomic and Molecular Physics,
Wuhan Institute of Physics and Mathematics, Innovation Academy of Precision Measurement Science and Technology, Chinese Academy of Sciences, Wuhan, 430071, China\\
$^{2}$ School of Physics, Zhengzhou University, Zhengzhou 450001, China \\
$^{3}$ School of Physics, University of the Chinese Academy of Sciences, Beijing 100049, China\\
$^{4}$ Research Center for Quantum Precision Measurement, Institute of Industry Technology, Guangzhou and Chinese Academy of Sciences, Guangzhou, 511458, China  }

\begin{abstract}
Quantum gates induced by geometric phases are intrinsically robust against noise due to their global properties of the evolution paths.
Compared to conventional nonadiabatic geometric quantum computation (NGQC), the recently proposed nonadiabatic noncyclic geometric quantum computation (NNGQC)
works in a faster fashion, while still remaining the robust feature of the geometric operations.
Here, we experimentally implement the NNGQC in a single trapped ultracold $^{40}$Ca$^{+}$ ion for verifying the noise-resilient and fast feature. By performing
unitary operations under imperfect conditions, we witness the advantages of the NNGQC with measured fidelities by quantum process tomography in
comparison with other two quantum gates by conventional NGQC and by straightforwardly dynamical evolution. Our results provide the first evidence confirming the possibility of accelerated quantum information processing with limited systematic errors even in the imperfect situation.
\end{abstract}
%\pacs{05.70.-a,37.10.Vz,03.67.-a}

\maketitle
Geometric quantum operations \cite{gqc1}, no matter adiabatic or nonadiabatic, are intrinsically noise-resilient due to their global properties of the evolution paths, which provide a promising paradigm for robust quantum information processing \cite{gqc3,gqc4,gqc5,gqc6,gqc7,gqc8,gqc9,gqc10,gqc11,gqc12,gqc13,gqc14,gqc15,gqc16,gqc17}. In particular, the nonadiabatic geometric
quantum computation (NGQC) \cite{ngqc1,ngqc2,ngqc3,ngqc4,ngqc5,ngqc6,ngqc7,ngqc8,ngqc9,ngqc10,ngqc11,ngqc12,ngqc13,ngqc14,ngqc15,ngqc16,ngqc17,ngqc18}, with less gating time than the adiabatic counterpart, could effectively protect against environment-induced decoherence. Various systems have already demonstrated such advantages with currently available techniques \cite{Engqc1,Engqc2,Engqc3,Engqc4,Engqc5,Engqc6,Engqc7,Engqc8,Engqc9,Engqc10,Engqc11,Engqc12,Engqc13}.

\begin{figure}[tbph]
\centering
\includegraphics[width=8.4cm,height=3.2 cm]{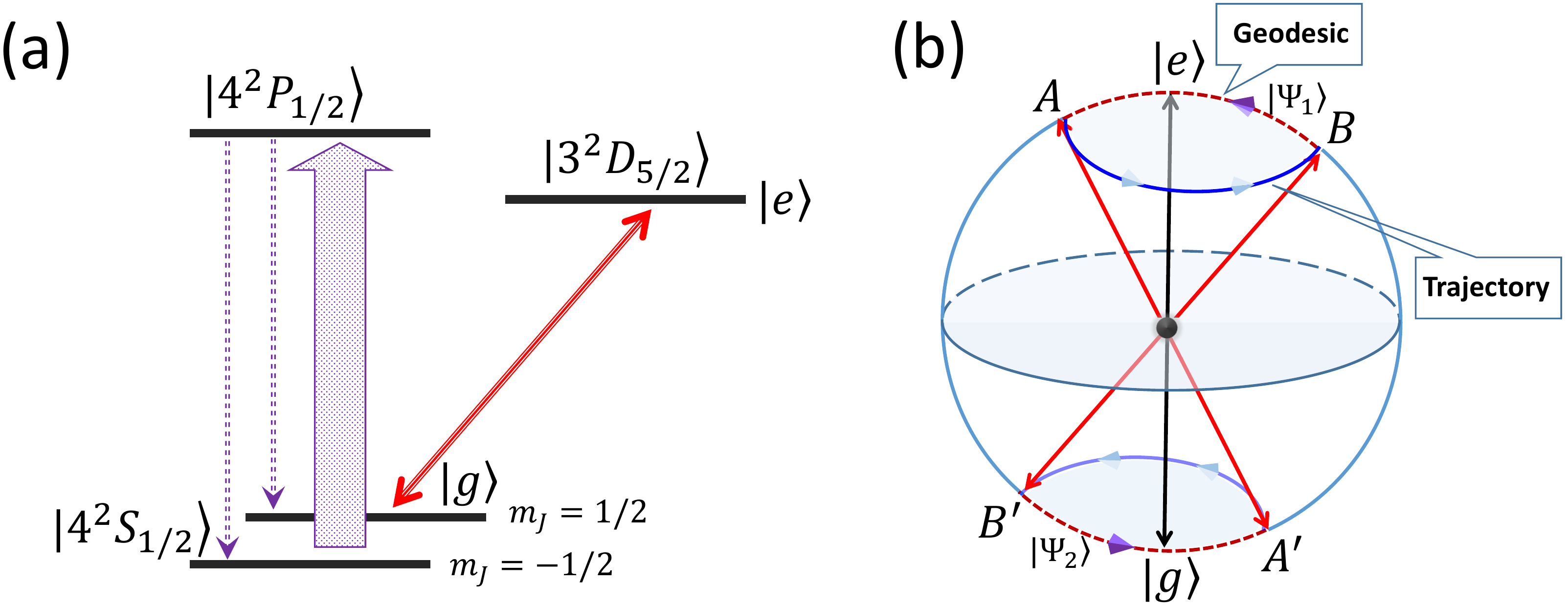}
\includegraphics[width=8.9cm,height=5.5 cm]{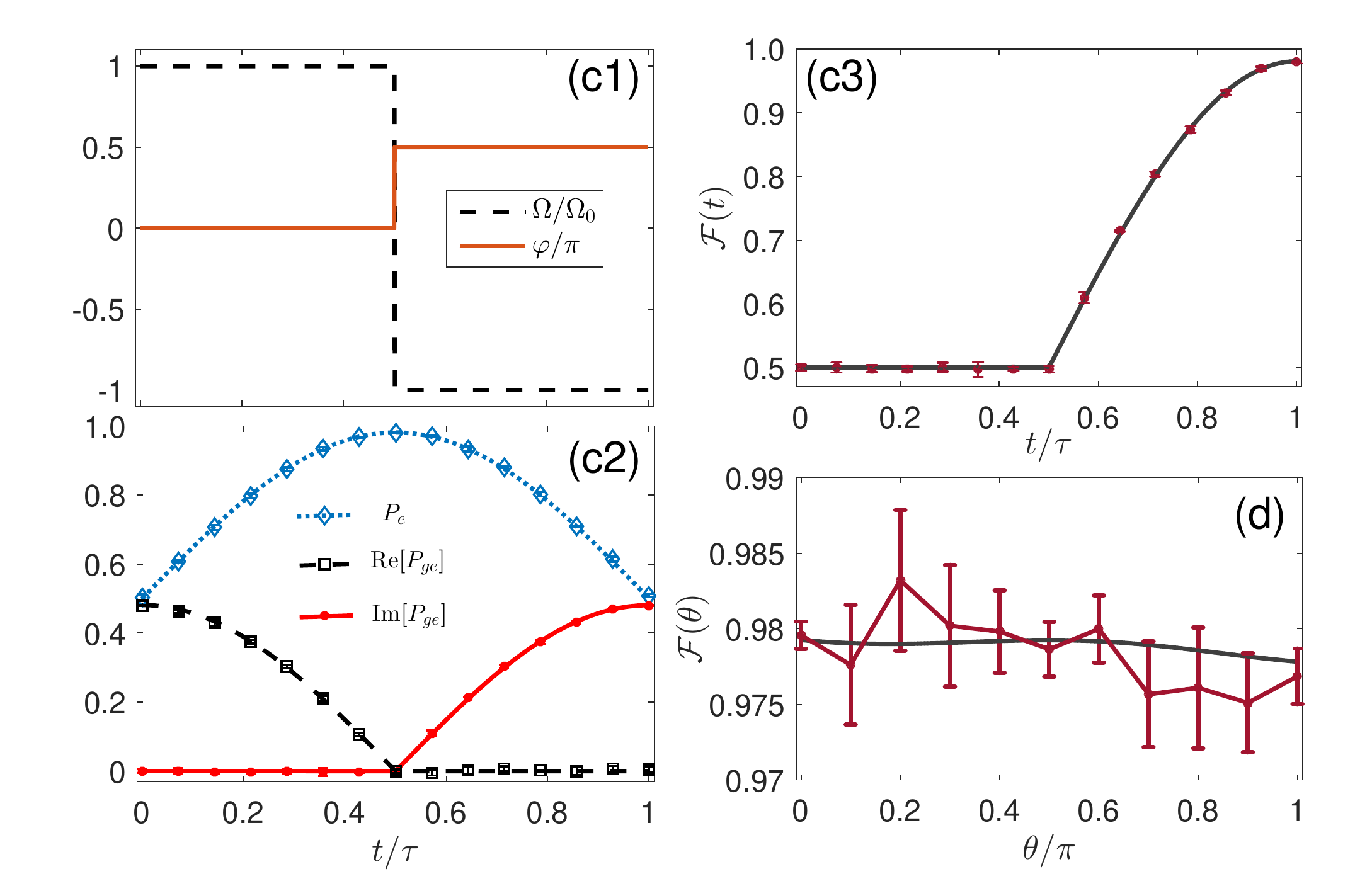}
\caption{(Color online) Single-qubit NNGQC. (a) Level scheme of the trapped $^{40}$Ca$^+$ ion, where the qubit is encoded in the ground state $|4^2S_{1/2},m_{J}=+1/2 \rangle$ and the metastable state $|3^2D_{5/2},m_{J}=+1/2 \rangle$. $|4^2P_{1/2} \rangle$ is the excited state employed for qubit readout. (b) Conceptual sketch of the evolution trajectories on the Bloch sphere for realizing single-qubit NNGQC gates, where the noncyclic geometric phase is relevant to the solid angle enclosed by the blue solid trajectory and the purple dashed geodesic between A (A') and B (B'), connecting the initial and final points. The realistic trajectories are determined by the gate-relevant parameters \cite{SM} and the geodesic connection is subject to the geodesic rule proposition \cite{new2}. The evolution trajectories of $|\Psi_{1}\rangle$ and $|\Psi_{2}\rangle$ are orthogonal and simultaneously reverse. (c1) Pulse sequences to realize the $U_1$ gate defined as $U_1\equiv[1+i \ -1-i;1-i \ 1-i]/2=(I+i\sigma_z)/2-i(\sigma_x+\sigma_y)/2$. (c2-c3) Experimental results of $U_1$ gate reaching the fidelity $0.9796(18)$ in the evolution started from the initial state $|\psi\rangle=(|g\rangle-i|e\rangle)/\sqrt{2}$, where $P_{e}$ is the population in state $|e\rangle$ and $P_{ge}$ is the off-diagonal term in the evolution matrix. (d) Observed fidelity of $U_1$ gate for different input states $|\psi(\theta)\rangle=\cos(\theta/2)|g\rangle-i\sin(\theta/2)|e\rangle$ with an average fidelity $0.9784(23)$. The error bars indicating the statistical standard deviation of the experimental data are obtained by 20000 measurements for each data point.}
\label{Fig1}
%1,0.9808;2,0.9789(4)
\end{figure}

The NGQC can be further accelerated if the time-optimal technology is involved to minimize gating time under the framework of cyclic evolution \cite{OPgqc2,OPgqc3,OPgqc4}. However, no matter with or without the time-optimal technology, the NGQC should satisfy the cyclic condition, which are more sensitive to decay and dephasing errors \cite{exc1,exc2,exc3,exc4} compared to the conventional dynamical quantum computation (DQC). In addition, the NGQC takes exactly the same amount of gating time
regardless of the large or small geometric rotation angle involved, which is subject to the cyclic condition, but impossibly changed by optimizing other parameters. Actually, the cyclic requirement is not always necessary in accomplishment of the NGQC, as indicated in a recent proposal \cite{PRR-2-043130}, in which the proposed nonadiabatic noncyclic geometric quantum computation (NNGQC) could reduce the gating time, while retaining the robust property of the geometric gates. In particular, compared to previously published noncyclic geometric models, e.g., \cite{new0,noncyclic1,noncyclic2}, the proposal using simpler pulse sequences could be executed with a faster pace and more relevance to current laboratory technique.
%The basic idea of the NNGQC originates from the non-Abelian nature of the state evolution in the phase space, which forms a geometric phase in the absence of cyclic condition.
Despite no experimental justification so far, the proposal presents us with hopes that practical quantum computation would be available in a way of high tolerance to errors and meanwhile high speed of gating.

Here we demonstrate a single-spin verification of this NNGQC implementation via experimental manipulation of a trapped-ion system.
Due to high-precision control, our execution on a qubit (i.e., a single spin) encoded in a single ultracold $^{40}$Ca$^{+}$ ion provides
an elaborate comparison of the NNGQC with others conventionally employed, such as the NGQC and DQC. As sketched in Fig. \ref{Fig1}(a), our qubit is encoded in the ground state $|4^2S_{1/2},m_{J}=+1/2\rangle$ (labeled as $|g\rangle$) and the metastable state $|3^2D_{5/2},m_{J}=+1/2 \rangle$ (labeled as $|e\rangle$), where m$_{J}$ is the magnetic quantum number. Although our investigation below only focuses on this qubit, to avoid thermal phonons yielding offsets of Rabi oscillation, we need to cool the ion to be within the Lamb-Dicke regime \cite{ion-1,ion-2,ion-3}.

Before specifying the experimental details, we first elucidate briefly the theory of constructing the single-qubit geometric gate using the basis states $\{|g\rangle,|e\rangle\}$.
For our purpose, the light-matter interaction is given by the Hamiltonian in units of $\hbar=1$ \cite{PRR-2-043130},
\begin{equation}
H=\frac{1}{2}\Omega(t)e^{i\varphi(t)}|e\rangle\langle g|+\rm H.C.,
\label{Eq1}
\end{equation}
which can be achieved in a single trapped ion by laser irradiation under carrier transitions. $\Omega(t)$ and $\varphi(t)$ are time dependent, representing the driving amplitude (i.e., Rabi frequency) and the phase of the laser, respectively. To understand the physics more clearly, we may map the system's states to two auxiliary states $|\Psi_{1,2}\rangle$.
In terms of the proposal \cite{PRR-2-043130}, realizing a NNGQC requires the amplitude and phase of the laser in time evolution to satisfy a couple of differential equations as below,
\begin{equation}
\Omega(t)=-\frac{\dot{\xi}}{\sin(\varphi+\eta)}, \quad \varphi(t)=-\eta+\arctan\frac{\dot{\xi}}{\dot{\eta}\tan\xi},
\label{Eq2}
\end{equation}
where $\xi$ and $\eta$ are time-dependent variables defined by $|\Psi_{1,2}\rangle$ spanning the space for constructing the evolution operator regarding the NNGQC \cite{SM}. To make the gates purely geometric, however, we have to erase the diagonal dynamical phase produced in the state evolution, implying $\varphi+\eta=-\pi/2$. In this case, the diagonal geometric phase is relevant to the solid angle enclosed by the evolution trajectory of $|\Psi_{1}\rangle$ or $|\Psi_{2}\rangle$ and the dashed geodesic connecting the initial and final points, see Fig. \ref{Fig1}(b). Compared to the NGQC that requires the evolution trajectories going through the two polars of the Bloch sphere, requirements for the NNGQC are much relaxed \cite{noncyclic1,new2}. As a result, the NNGQC could take less time than that of the NGQC for the gate implementation. For example, in U$_{1}$ gate implementation, the NNGQC takes half of time compared to the NGQC, and less than half of time than DQC \cite{SM}.

For a universal set of single-qubit NNGQC gates, we may consider the following parameter set
\begin{equation}
\xi=\Omega_0t-\xi_0, \quad \eta=\phi_1\varepsilon(t)+\phi_0 ,
\label{Eq3}
\end{equation}
where $\Omega_0,\phi_0,\phi_1$ and $\xi_0$ are constants dependent on the designed phase gate, and $\varepsilon(t)$ is the step function dividing the evolution with duration $\tau$ into two parts: $\varepsilon(t)=0$ for $t\in [0,\xi_0/\Omega_0]$ and $\varepsilon(t)=1$ in the case of $t\in [\xi_0/\Omega_0,\tau]$. These constants constitute four parameters $\eta_{\pm}$ and $\xi_{\pm}$, with $\eta_+=\phi_1+2\phi_0$, $\eta_-=\phi_1$, $\xi_+=\Omega_0\tau$, and $\xi_-=\Omega_0\tau-2\xi_0$, which are relevant to the rotating angles $\theta$, $\alpha$ and $\beta$ regarding the NNGQC, as indicated below. The total evolution operator is given by
\begin{equation}
U(\theta,\alpha,\beta)=Z_{\beta}X_{\theta}Z_{\alpha},
\label{Eq4}
\end{equation}
where the rotation operators regarding $z$ and $x$ axes of the Bloch sphere are defined as $Z_{\beta}\equiv e^{-i\beta\sigma_z/2}$ and $X_{\theta}\equiv e^{-i\theta\sigma_x/2}$. The angles are given by $\theta=2\sin^{-1}\sqrt{\sin^{2}\gamma \sin^{2}(\xi_+/2)+\cos^{2}\gamma \sin^{2}(\xi_-/2)}$, $\alpha=-\tan^{-1}\frac{\tan\gamma\cos(\xi_+/2)}{\cos(\xi_-/2)}-\tan^{-1}\frac{\tan\gamma\sin(\xi_+/2)}{\sin(\xi_-/2)}+\frac{\eta_--\eta_+-\pi}{2}$,
and $\beta=-\tan^{-1}\frac{\tan\gamma\cos(\xi_+/2)}{\cos(\xi_-/2)}+\tan^{-1}\frac{\tan\gamma\sin(\xi_+/2)}{\sin(\xi_-/2)}+\frac{\eta_-+\eta_++\pi}{2}$ \cite{SM}, where $\gamma=\int_{0}^{\tau}i\langle\Psi_{1}(t)|(d/dt)|\Psi_{1}(t)\rangle=\phi_{1}/2$ is relevant to the NNGQC geometric phase.
To execute $U_1$ (Hadamard) gate, we choose $\phi_0=-\pi/2 ~(0)$, $\phi_1=\pi/2 ~(\pi/2)$, $\xi_0=\pi/2 ~(\pi/2)$ and $\tau=\pi/\Omega_0 ~(3\pi/2\Omega_0)$, as proposed in \cite{PRR-2-043130}. These are exactly the gates we will execute experimentally below.

\begin{figure}[tbhp]
\centering
\includegraphics[width=8.4cm,height=1.2 cm]{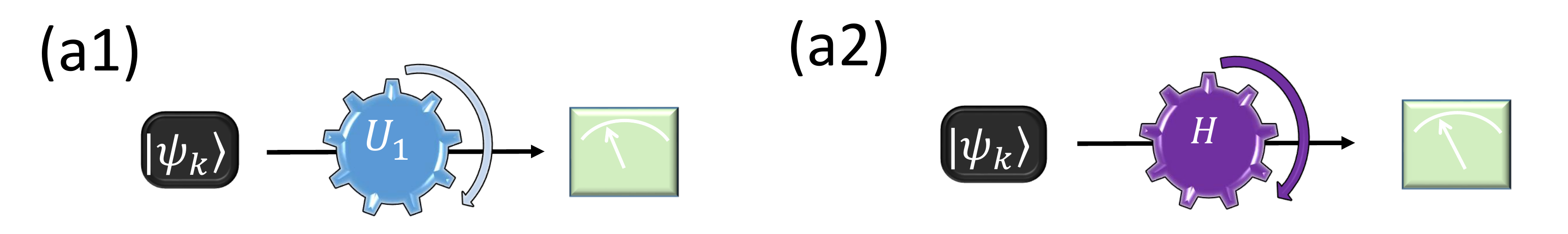}
\includegraphics[width=8.4cm,height=5.5 cm]{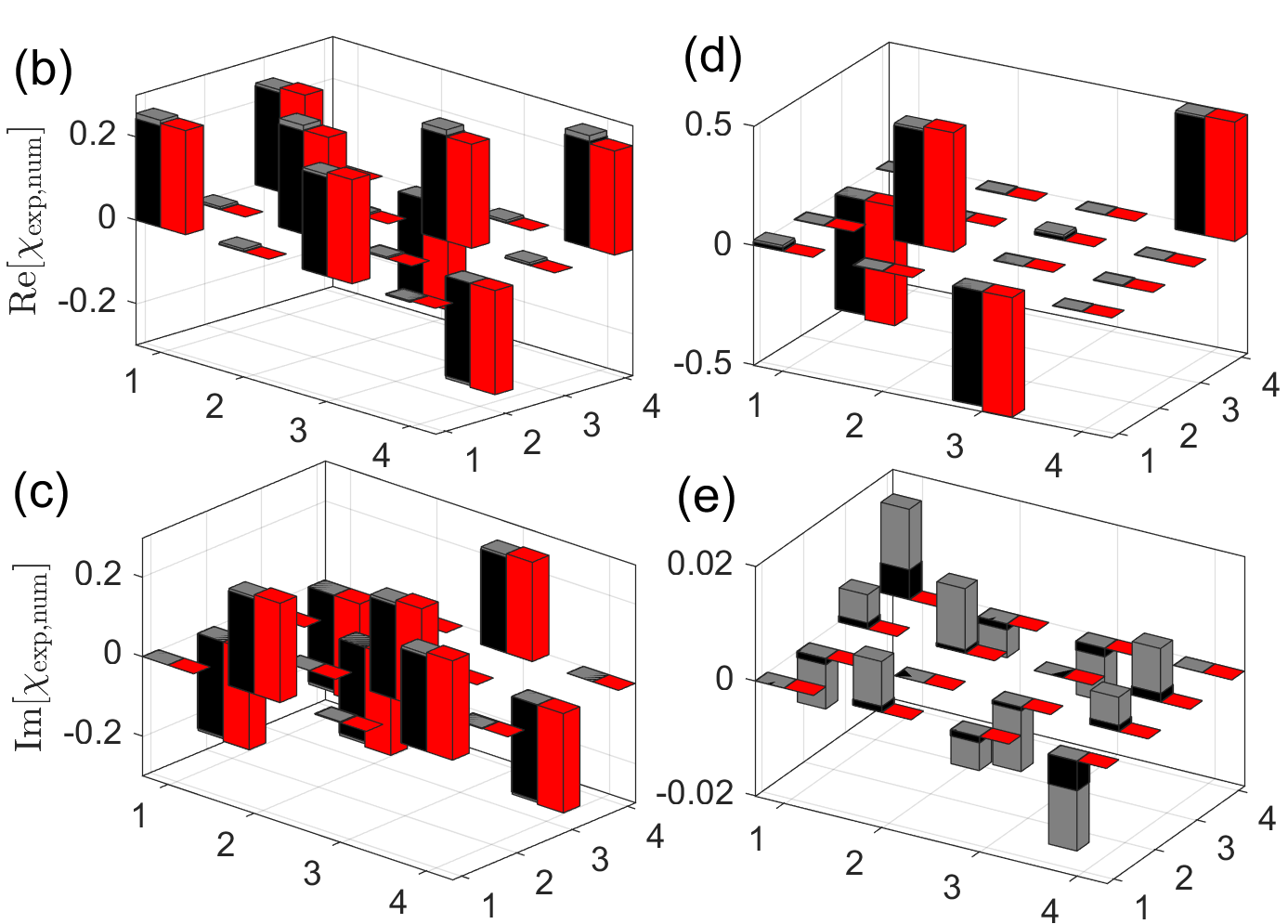}
\caption{(Color online)  Experimental implementation of a single-qubit NNGQC for (a1) $U_1$ gate as defined in Fig. 1 and (a2) Hadamard gate $H$, where the input states are eigenvectors of the operator set $\{ I,\sigma_x,-i\sigma_y,\sigma_z\}$ regarding the basis states $\{|g\rangle,|e\rangle\}$. (b-d) The experimental (black) and numerical (red) bar charts for the real and imaginary parts of the process matrix $\chi_{\rm exp, num}$, respectively, where the grey parts above the experimental (black) bars denote the errors of the experimental data acquired by 20000 measurements. The numbers labeled in the axes represent four input states, i.e., $|g\rangle$, $|e\rangle$, $|+\rangle=(|g\rangle+|e\rangle)/\sqrt{2}$, and $|-\rangle=(|g\rangle-i|e\rangle)/\sqrt{2}$.
(b,c) $U_1$ gate with the process fidelity of $0.9658(34)$; (d,e) Hadamard gate with  the process fidelity of $0.9645(27)$.}
\label{Fig2}
%theoretical value,U1:0.9681;H:0.9667
\end{figure}

\begin{figure*}[bhtp]
\centering
\includegraphics[width=15.6cm,height=1.6 cm]{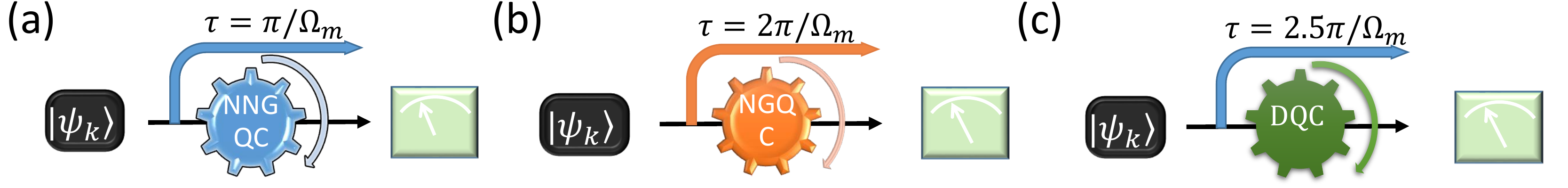}
\includegraphics[width=16.0cm,height=5.2 cm]{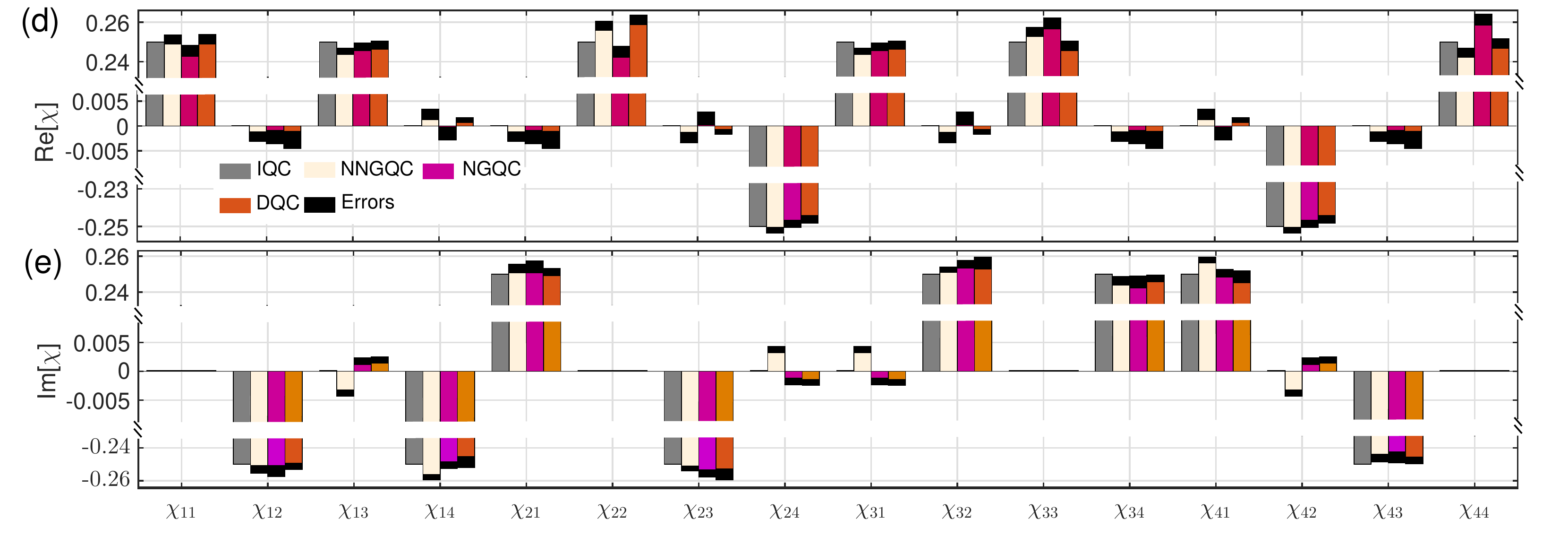}
\caption{(Color online) Comparison of NNGQC with NGQC and DQC. (a-c) Schematic for processing a $U_1$ gate using the three gate schemes with $\Omega_{m}$ the constant Rabi frequency as explained in \cite{SM}. (d,e) Histogram charts for the real and imaginary parts of $\chi$ under different quantum gate processes, where focusing on the data around $0$ and $0.25$ is to display the difference among different gate schemes and distinguish small errors of the experimental data. To distinguish the slight differences, we increase the measurement times to 50000 for each data. For a full comparison, we also plot the ideal $U_1$ gate here, labelled as IQC.}
%The process fidelities of the $U_1$ gate, after deducting the deviation caused by the population leakage, are 0.9978(16), 0.9936(22) and 0.9918(16) for NNGQG, NGQG and DQG, respectively. }
\label{Fig3}
%theoretical value,ideal: 1;NNGQC:0.9959;NGQC:0.9929;DQC:0.9915
\end{figure*}

In our experiment, the $^{40}$Ca$^{+}$ ion is confined in a linear Paul trap with axial frequency $\omega_z/2\pi=1.1$ MHz and radial frequency $\omega_r/2\pi=1.6$ MHz.
We define a quantization axis along the axial direction by a magnetic field of approximately 3.4 Gauss at the center of the trap, and
%The transitions between $|g\rangle$ and $|e\rangle$ are connected via a quadrupole transition at a wavelength of 729 nm.
manipulate the qubit by a ultra-stable narrow linewidth 729-nm laser at an angle of 60 degrees to the quantization axis with the Lamb-Dicke parameter $\sim$0.1152.
The 729-nm laser is controlled by a double pass acousto-optic modulator. We have the field programable gate array to control a direct digital synthesizer
for the frequency sources of the acousto-optic modulator, which provides the phase and frequency control of the 729-nm
laser during the experimental operations. Before starting our experiment, we have cooled the ion, from the three dimensions, down to near the ground state of the
vibrational modes, yielding the final average phonon number $ \bar{n}< $1. In addition to other noises caused by the magnetic and electric field fluctuation,
the qubit suffers from the dephasing of $\Gamma=0.81(11)$ kHz, as measured by Ramsey sequences. However,
since the average Rabi frequency regarding the 729-nm laser irradiation is $\Omega_0/2\pi=67.9(2)$ kHz and the decay rate ($<$ 1 Hz) is negligible,
we estimate that the dephasing-induced infidelity is about 0.1$\%$ \cite{SM}, which
should be carefully treated in order to distinguish the measured fidelities of the three different gate schemes as elucidated below.

To deliberate on the noise-resilient characteristic of the NNGQC, we also introduce in our operations some other imperfection, respectively, regarding the initial state preparation, the Rabi frequency and the resonance frequency, which are the primary error sources in most trapped-ion experiments \cite{RMP-75-281}. We first carry out the $U_1$ gate by the NNGQC scheme in the presence of imperfect initial state preparation, where the fidelity is defined by $\mathcal{F}=\text{Tr}[\rho_{\rm exp} \rho_{\rm id}]$ with 'exp' and 'id' denoting the experimental and ideal states. The time sequences of the Rabi frequency and the laser phase for achieving the NNGQC are designed in Fig. \ref{Fig1}(c1), whose discontinuous features are also reflected in Fig. \ref{Fig1}(c2,c3) as the cusps found in the observed fidelities and the off-diagonal terms $P_{ge}$ regrading the output matrix of the state. Despite the small infidelity as observed in Fig. \ref{Fig1}(d), the experimental results coincide well with the numerical results, whose average fidelity is $\mathcal{F}=0.9789(4)$.

To further characterize the performance of the NNGQC in this case, we employ quantum process tomography (QPT) \cite{jmo-44-2455,Nie-Chu} for a full measurement of the experimental process matrices $\chi_{\rm exp}$. With the specific input state set $|\psi_k\rangle \in \{|g\rangle, \ |e\rangle$, $ |+\rangle=(|g\rangle+|e\rangle)/\sqrt{2}, \ |-\rangle=(|g\rangle-i|e\rangle)/\sqrt{2} \}$, we have carried out the $U_1$ and Hadamard (i.e., $H$) gates under the pulse sequences as plotted in Fig. \ref{Fig2}(a1, a2). We evaluate their process fidelities by $\mathcal{F}=\text{Tr}[\chi_{\rm exp}\chi_{\rm id}]$, which are $0.9658(34)$ for $U_1$ gate as in Fig. \ref{Fig2}(b,c) and $0.9645(27)$ for $H$ gate, see Fig. \ref{Fig2}(d,e). This observation is consistent with the corresponding values of our numerically predicted fidelities, i.e., $0.9681$ and $0.9667$.

As predicted in \cite{PRR-2-043130}, the NNGQC works faster and also owns a higher fidelity than the NGQC and DQC in accomplishing the same quantum gating tasks. The higher gating speed is easily understandable from Fig. \ref{Fig3}(a,b,c) that the NNGQC only takes half of the gating time compared to the NGQC, and takes much less gating time than the DQC \cite{SM}. However, distinguishing the fidelity difference among the three gate schemes, which is theoretically predicted to be less than $0.3\%$, is experimentally challenging \cite{SM}. To display the fidelity superiority of the NNGQC, we have increased the measurement times to 50000 for each data point, yielding better fidelities of the gate performance, i.e., $0.9700(20)$, $0.9660(18)$ and $0.9642(17)$ for NNGQC, NGQC and DQC, respectively. Correspondingly, we have the numerically calculated fidelities to be $0.9681$, $0.9653$ and $0.9639$. These values show the similar trend that both the geometric gates work with evidently higher fidelities than the DQC, where the NNGQC also has a better performance than the NGQC even in the presence of imperfection. After calibrating the data over the imperfection in the initial state preparation, we have updated the process matrix elements as shown in Fig. \ref{Fig3}(d,e). The corresponding fidelities of $0.9978(16)$, $0.9936(22)$ and $0.9918(16)$ are still consistent with the trend of the calibrated numerical results $0.9959$, $0.9929$ and $0.9915$. This implies that the NNGQC works better in both cases with and without the perfect preparation of the initial state. It also indicates that our calibration is beneficial to improve the fidelities of the gates, but without essentially changing the main features of the gates in the comparison.

\begin{figure}[tbph]
\centering
\includegraphics[width=8.2cm,height=6.8 cm]{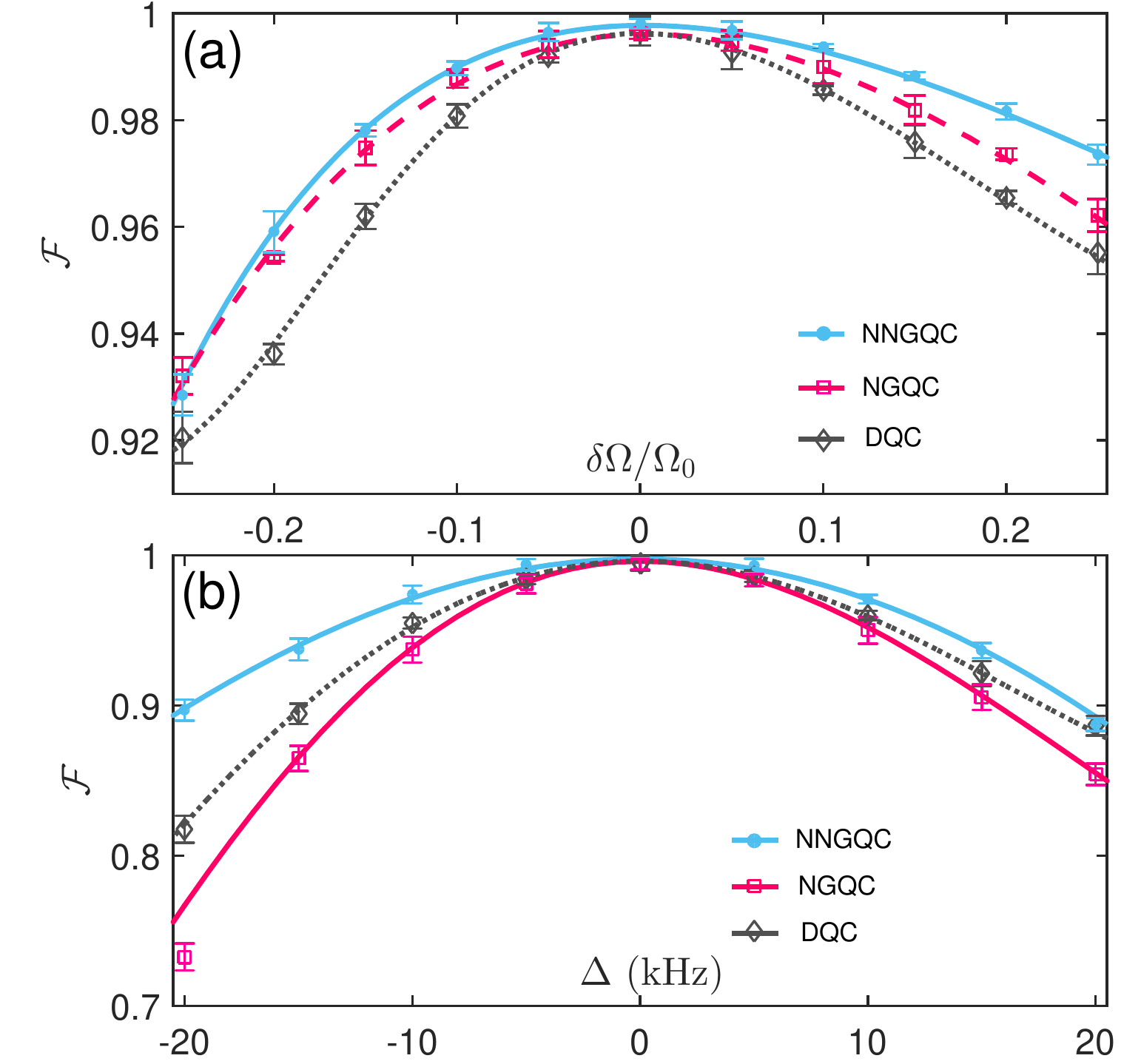}
\caption{(Color online) Comparison of the robustness against systematic errors based on the experimental fidelities of the output states after undertaking the $U_1$ gates carried out, respectively, by the NNGQC, NGQC and DQC schemes for the input state $|g\rangle$. (a) For the Rabi frequency error $\delta\Omega$ of the laser and (b) for the resonance frequency error $\Delta$ of the qubit. The experimental results are consistent with the numerical simulation results. The error bars indicating the statistical standard deviation of experimental data are obtained by 20000 measurements for each data point.}
\label{Fig4}
%0.1Rabi fluctuation:errors,NNGQC:T:0.9956(65),E:0.9981(66);NGQC:T:0.9929(77),E:0.9961(75);DQC:T:0.9915(115),E:0.9967(122);
%10kHz Detuning fluctuation:errors,NNGQC:T:0.9956(259),E:1.0001(273);NGQC:T:0.9929(478),E:0.9941(499);DQC:T:0.9915(377),E:0.9947(354);
\end{figure}

In Fig. \ref{Fig4}, we justify the advantages of the NNGQC experiencing two other errors. To deliberate on this robustness, we have explored a wide range of the deviation from the correct values. We can find from the measurements that the exact resonance frequency is more important than the exact Rabi frequency  since a small deviation from the required resonance frequency affects the fidelity more seriously than the counterpart of the Rabi frequency. Nevertheless, the NNGQC owns a significant superiority over the other two gating schemes in resisting both of the errors. For example, for the frequency deviation of $|\Delta|/2\pi\leq 10$ kHz, which is actually a relatively large error compared to the Rabi frequency $\Omega_0/2\pi=67.9$ kHz, the NNGQC could still reach the fidelity larger than 97$\%$, much more robust than the NGQC and the DQC. In fact, even for a large error, e.g., $\pm 20\%$ deviation of $\Omega$ or $\pm$20 kHz of detuning, the NNGQC could still perform better than the other two gates in the same situation.

For a more quantitative comparison, we assume the fluctuation errors to obey the Gaussian distribution, i.e., $p(\epsilon)=\frac{1}{\sqrt{2\pi}\sigma}\exp(-\frac{\epsilon^2}{2\sigma^2})$ with $\sigma$ representing the standard deviation of the exemplified errors. We evaluate these Gaussian standard deviations in Table \ref{Table1}, which fit well the trend of the numerically calculated values and also indicate the prominent advantage of the NNGQC in the error resistance regarding the imperfect Rabi frequency and the inaccurate resonance frequency.

\begin{table}
\caption{Gaussian standard deviation (GSD) comparison of the NNGQC with the NGQC and DQC regarding the systematic errors from Rabi frequency and resonance frequency, where the GSD is estimated from the Gaussian distribution.}
\centering
\begin{tabular}{|c|c|c|c|c|c|c|c|c|c|c|c|c|c|c|c|c|c|}
\hline
\multirow{2}{*}{Parameters}& \multirow{2}{*}{GSD} & \multicolumn{2}{c|}{NNGQC} & \multicolumn{2}{c|}{NGQC} & \multicolumn{2}{c|}{DQC}  \\
\cline{3-8}
& & Exp  &  Num   & Exp  &  Num & Exp  &  Num \\
\hline
\multirow{2}{*}{$\sigma_{\Omega}/\Omega_0$} &0.1 & 0.66\% & 0.65\%& 0.75\%& 0.77\% & 1.22\% &1.15\% \\
\cline{2-8}
& 0.2 & 1.59\% & 1.36\%& 1.74\%& 1.55\% & 2.27\% &2.04\% \\
\hline
\multirow{2}{*}{$\sigma_{\Delta}$ (kHz)} &5 & 0.68\% & 0.67\%& 1.22\%& 1.28\% & 0.94\% &1.02\% \\
\cline{2-8}
& 10 &  2.30\%& 2.38\%& 4.13\% &4.43\% &3.13\% &3.50\% \\
\hline
\end{tabular}
\label{Table1}
\end{table}

In summary, our experiment has provided the first single-spin evidence confirming the fast and noise-resilient feature of the NNGQC in qubit manipulation. In particular, our high-precision control at the fundamental level of a single spin has displayed clearly the higher fidelity of the gate performance accomplished by the NNGQC than by others under some imperfect conditions. In contrast to the theoretical prediction that the NNGQC is superior to NGQC under the same environmental decoherence due to shorter gating time, our experimental observation has witnessed the advantages of the NNGQC regarding the factors more than decoherence influence, which would stimulate further study both theoretically and experimentally. Despite the very small differences presented here in the comparison, the higher fidelity, the higher tolerance to imperfection and the higher speed of the NNGQC gates would be of a great impact in multi-sequences of gate operations in a practical quantum information processing.

Considering the two-qubit controlled gates, which are the essential element in constituting universal quantum computation, we have found that the two-qubit controlled phase gates can be executed from an effective Hamiltonian very similar to Eq. (1) \cite{SM}. Thus, two-qubit gates of the NNGQC can be in principle carried out with the similar operations to the single-qubit gate as demonstrated above. This implies that our experimental verification here has actually witnessed the advantages of both the single-qubit and two-qubit operations required by the NNGQC \cite{explain}. In particular, for the two-qubit controlled gates, no matter whether they are achieved by direct coupling of the two levels or by a Raman process, the characteristics of the NNGQC would help for gating operations with higher-speed, higher-fidelity and lower error fashion, under the same conditions or parameters, in comparison with the same gating performance by other ways.
%Moreover, even for a quantum computation employing other two-qubit gate schemes without involving geometric feature, the present work investigating the NNGQC in the presence of finite phonons
%is still very useful. In combination with the M\o{}lmer-S\o{}renson entangling gate \cite{SMgate}, our effort actually indicates the availability of a practical quantum computation of 'hot' trapped ions without the stringent prerequisite of the ultracold states.

This work was supported by Key Research $\&$ Development Project of Guangdong Province under Grant No. 2020B0303300001, by National Key Research $\&$ Development Program of China under grant No. 2017YFA0304503, by National Natural Science Foundation of China under Grant Nos. 12074346, 12074390, 11835011, 11804375, 11804308, 91421111, 11734018, by Natural Science Foundation of Henan Province under Grant No 202300410481 and by K. C. Wong Education Foundation (GJTD-2019-15).\\

J. W. Z, L. L. Y. and J. C. Li contributed equally to this work.

\myemptypage
\pagebreak
%\widetext
	\begin{center}
	\textbf{\large Supplementary material for "Single-atom verification of the noise-resilient and fast characteristics of universal nonadiabatic noncyclic geometric quantum gates"}
\end{center}

\setcounter{equation}{0}
\setcounter{figure}{0}
\setcounter{table}{0}
\setcounter{page}{1}
\makeatletter
\renewcommand{\theequation}{S\arabic{equation}}
\renewcommand{\thefigure}{S\arabic{figure}}
\renewcommand{\bibnumfmt}[1]{[S#1]}
\renewcommand{\citenumfont}[1]{S#1}
%%%%%%%%%% Prefix a "S" to all equations, figures, tables and reset the counter %%%%%%%%%%

\section{Effect of dephasing on the fidelity of the gate} \label{Dephasing}

\begin{figure}[tbph]
	\centering
	\includegraphics[width=9 cm,height=8.7 cm]{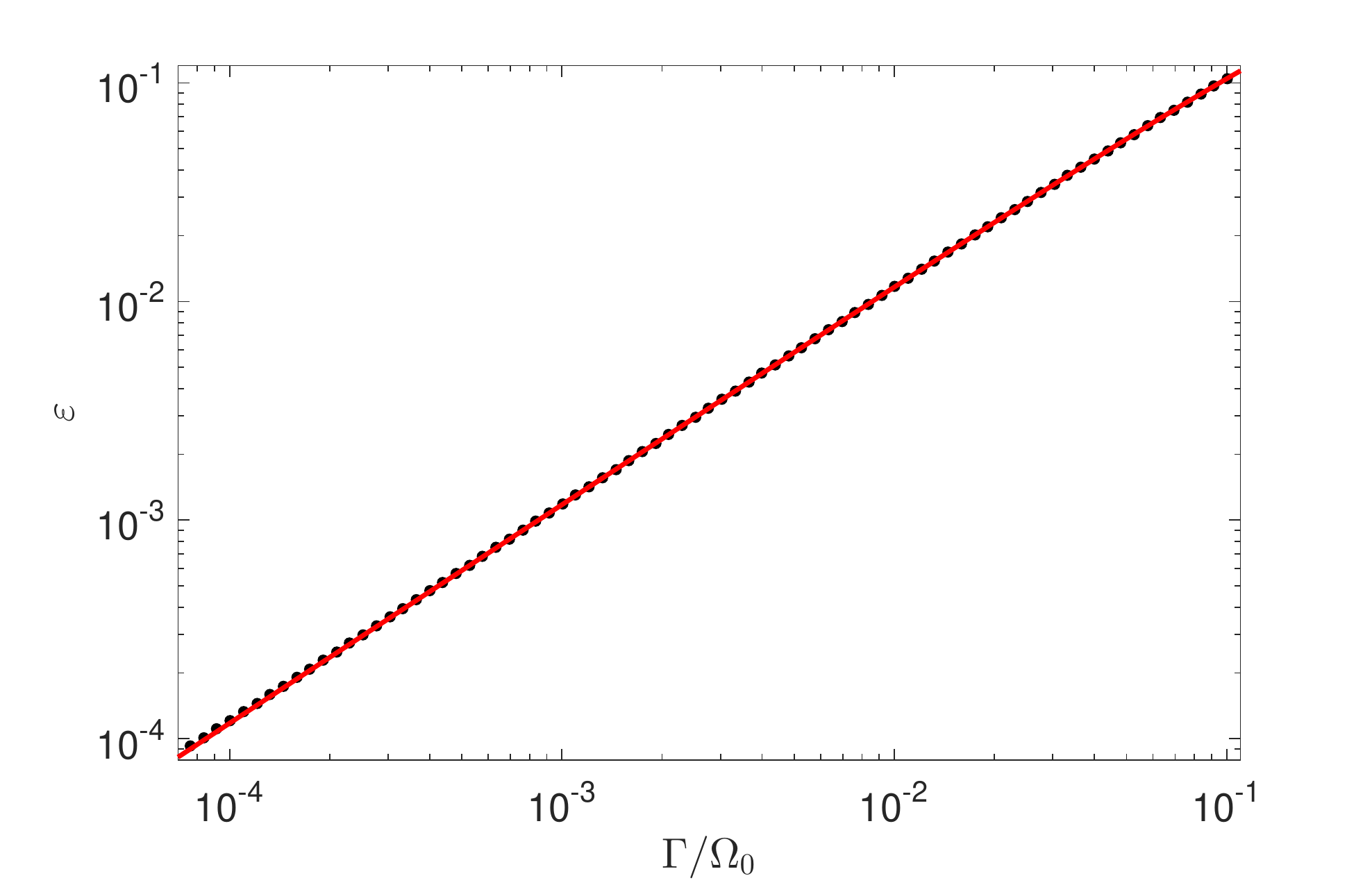}
	\caption{Infidelity as a function of dephasing, where the parameters are chosen as $\xi_0=\pi/2,\phi_0=-\pi/2,\phi_1=\pi/2$, and $\tau=\pi/\Omega_0$ with the Rabi frequency $\Omega_0/2\pi=0.1$ MHz to construct a $U_1$ geometric gate. The fitting curve is plotted by the function of $\varepsilon=a(e^{-b\Gamma/\Omega_0}-1)$ with the fitting parameters of $a=-0.4813$ and $b=2.447$, respectively.}
	\label{Fig2s1}
\end{figure}

Time evolution of the system's state under the dephasing effect can be described by Master equation as below,
\begin{equation}
\dot{\rho}=-i[H,\rho]+\frac{\Gamma}{2}(\sigma_z\rho\sigma_z-\rho),
\end{equation}
where $\Gamma$ denotes the dephasing rate of the system. We define the infidelity caused by the dephasing as
\begin{equation}
\varepsilon=1-\mathcal{F},
\end{equation}
where the fidelity of the quantum gate is defined as $\mathcal{F}=\text{Tr}[\rho\rho_{\rm ideal}]$ with $\rho$ and $\rho_{\rm ideal}$ corresponding to the real and ideal final states after accomplishing the quantum gate. Considering the weak dephasing in our experiment, we only plot the cases of $\Gamma\ll \Omega_0$ in Fig. \ref{Fig2s1},
where to obtain an infidelity smaller than $10^{-2}$, we require $\Gamma/\Omega_0< 8.6 \times 10^{-3}$. In our case, we have experimentally measured the Rabi frequency of $\Omega/2\pi=67.9(2)$ kHz and the dephasing rate of $\Gamma=0.81(11)$ kHz, implying that the infidelity is only 0.1$\%$.

%the deviation will increase linearly as increase of $\Gamma$ in the weak dephasing region, i.e.,. To obtain a higher fidelity of gate, it needs a small dephasing rate, for example, in Fig \ref{Fig2s1}, which means that the dephasing time should be longer than 185 $\mu$s for a Rabi frequency $\Omega_0/2\pi=0.1$ MHz and longer than 1.9 ms for obtaining a fidelity larger than 0.999. Furthermore, since the deviation caused by the dephasing are constant in a decided evolution time, it can be corrected by comparing with the accurate numerical results.

\section{Errors caused by quantum projection measurement}

Assuming the success probability of the measurement to be $p$, we have the error caused by quantum projection measurement given by $\epsilon=\sqrt{p(1-p)/N}$ with $N$ the times of the projection measurement. Since the phase gate of a single qubit needs three different measurements of the elements of the density matrix $\rho$, the gate error $\epsilon_g$ is $\sqrt{3}$ times larger than a single projection measurement error. However, the projection measurement errors of the gate are intrinsic and random, which are impossibly calibrated. Nevertheless, due to the fact of $\epsilon_g\propto 1/\sqrt{N}$, increasing the measurements is an effective way to suppress this error.
%with the variance $\epsilon_g$ and the mean value $0$ for the given measurement times $N$. Thus, it cannot correct the projection measurement errors by the numerical calculation.

%\begin{figure}[tbph]
%\centering
%\includegraphics[width=8.0cm,height=5.7 cm]{Fig2s2.eps}
%\caption{The numerical projection error of phase gate as the increase of the quantum projection measurement times where the errors are obtained by 1000 repetitions of projection measurement process. The dephasing rate is set as $\Gamma/\Omega_0=3.2\times 10^{-3}$ and the other parameters are chosen same as Fig. \ref{Fig2s1}. The fitting function of fitting curve is selected as $\epsilon_g=g_e/\sqrt{N}$ with the fitting parameter $g_e=0.5$.}
%\label{Fig2s2}
%\end{figure}

In order to distinguish the small difference among the measured fidelities of the three different gates as mentioned in the main text, we need to suppress the projection measurement errors. Due to the difference being smaller than 0.5$\%$, we have made 50000 measurements for each data point in Fig. 3 of the main text.

\section{Errors caused by imperfect preparation of the initial state}

In our experiment, the qubit is initialized by the 397 nm laser, which prepares the initial state in the ground state $S_{1/2}$ with the population $p=0.9628(1)$ with imperfection $dp=1-p=0.0372(1)$. Thus, when we exclude this imperfection, we calibrate the experimental data  by the renormalized factor $p$.

%Thus, it needs a suitable projection measurement times for a given mean values difference. For example, for two different protocols to construct a phase gate, we obtain the fidelities $\mathcal{F}_1=0.99$ and $\mathcal{F}_2=0.995$ that the difference between them is $\delta\mathcal{F}=0.005$, thus, the errors for them should satisfy $\epsilon_g<0.0025$. Using the numerical results in Fig. \ref{Fig2s2}, it needs $N>40000$. In our text, the gate fidelity differences for NNGQC and NGQC are in this rang, thus, selecting the measurement time larger than $40000$ is suitable for the experimental implement. On the other hand, other errors may also be enhanced in the process which thus needs a larger measurement times. Therefore, we consider the error difference between the difference quantum gate protocols by using $N=50000$ in the Fig. 3 of main text while in other figures we adopt $N=20000$ due to the difference in other figures is about $\delta\mathcal{F}=0.01$.

\section{Some details for the gates by nonadiabatic noncyclic geometric quantum computation}

\begin{figure}[tbph]
	\centering
	\includegraphics[width=8.8cm,height=5.1 cm]{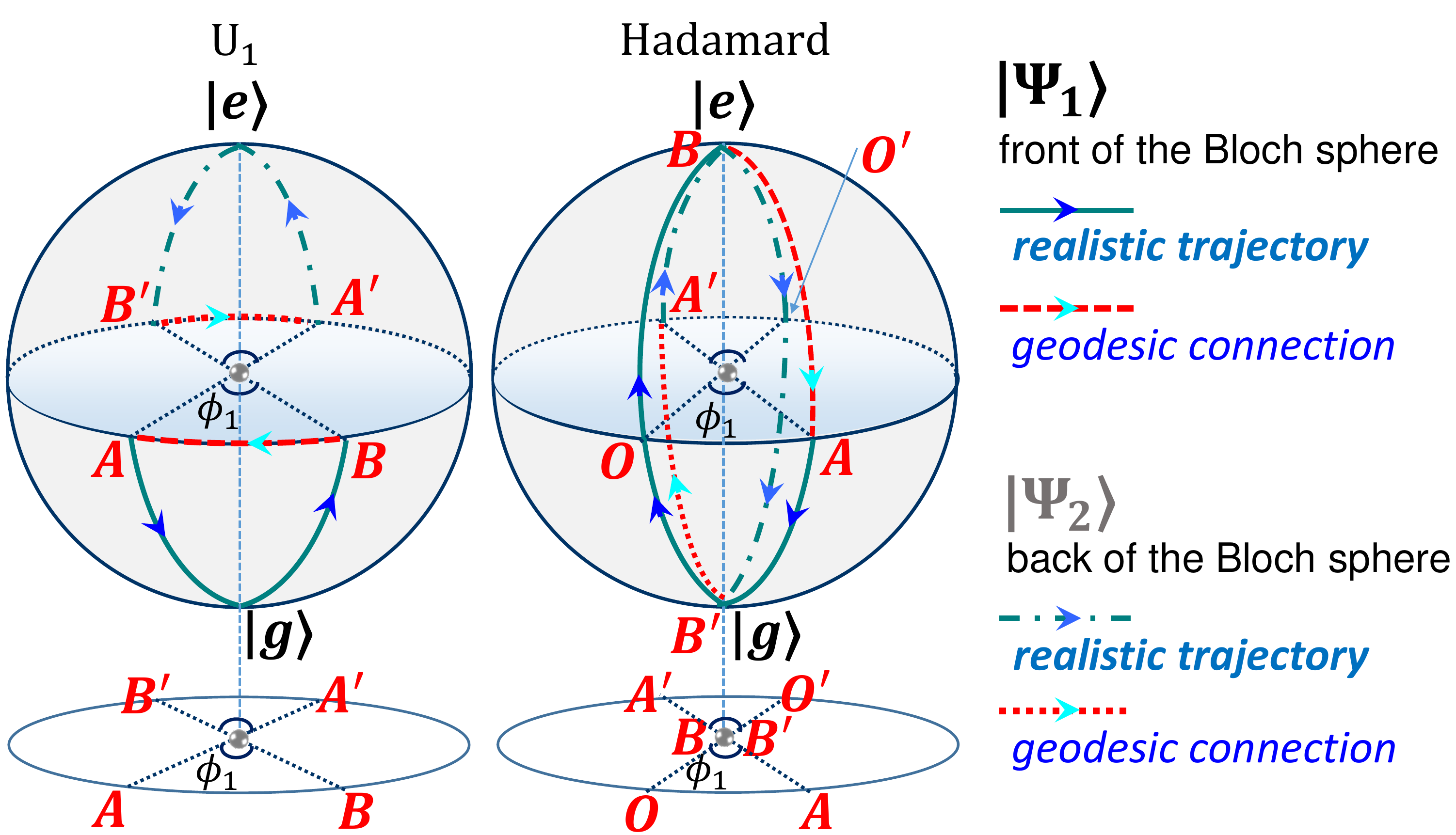}
	\caption{Illustrations of the trajectory and geodesic connection in the Bloch sphere for the NNGQC $U_{1}$ and Hadamard gates. The evolution operator for the NNGQC is $U(\tau,0)=e^{i\gamma}|\Psi_{1}(\tau)\rangle\langle\Psi_{1}(0)| + e^{-i\gamma}|\Psi_{2}(\tau)\rangle\langle\Psi_{2}(0)|$. With the parameters from the same group, $|\Psi_1\rangle$ and $|\Psi_2\rangle$ evolve simultaneously along the trajectories plotted here. In the case of the dynamical phase being zero, the geometric phase is generally given by $\gamma + \arg\langle\Psi_{1}(0)|\Psi_{1}(\tau)\rangle$ \cite{new,ae2003}, where for the cyclic case (i.e., $|\Psi_1(\tau)\rangle=|\Psi_1(0)\rangle$) the geometric phase is simply $\gamma$.
		%, while for the NNGQC case the geometric phase is $\gamma + \arg\langle\Psi_{1}(0)|\Psi_{1}(\tau)\rangle$.
		The beginning and ending points of the trajectory for $|\Psi_{1}\rangle$($|\Psi_2\rangle$) are $A(A')$ and $B(B')$, respectively. It should be noted that since the dynamical phase is zero throughout the evolution in our scheme, the realistic trajectories also evolve along the geodesic lines, which is different from in Refs.~\cite{ae2003,sciadvgeodesic} involving non-zero dynamical phase during the evolution process.}
	\label{figbloch} %Thus, it may be better if the trajectory and geodesic for both of auxiliary states are shown.
\end{figure}

The rotation angles regarding the NNGQC, as mentioned in the main text, are of the general forms as follows,
\begin{eqnarray}
\theta&=&2\sin^{-1}(\sqrt{\Lambda_{2\gamma,\xi_+}^2+\nu_{\xi_-,2\gamma}^2}), \notag\\
\alpha&=&-\tan^{-1}\frac{\nu_{2\gamma,\xi_+}}{\mu_{2\gamma,\xi_-}}-\tan^{-1}\frac{\Lambda_{2\gamma,\xi_+}}{\nu_{\xi_-,2\gamma}}+\frac{\eta_--\eta_+-\pi}{2}, \notag\\
\beta&=&-\tan^{-1}\frac{\nu_{2\gamma,\xi_+}}{\mu_{2\gamma,\xi_-}}+\tan^{-1}\frac{\Lambda_{2\gamma,\xi_+}}{\nu_{\xi_-,2\gamma}}+\frac{\eta_-+\eta_++\pi}{2}, \notag
\end{eqnarray}
where $\gamma$ and $\xi_{\pm}$ are angles as specified below, and other corresponding parameters are defined as $\mu_{k,j}\equiv\cos\frac{k}{2}\cos\frac{j}{2}$, $\nu_{k,j}\equiv\sin\frac{k}{2}\cos\frac{j}{2}$,  $\Lambda_{k,j}\equiv\sin\frac{k}{2}\sin\frac{j}{2}$, $\xi_{\pm}=\xi(\tau)\pm\xi(0)$, and $\eta_{\pm}=\eta(\tau)\pm\eta(0)$.
To see the realistic trajectory and geodesic connection, we have to introduce two auxiliary basis states $|\Psi_1\rangle=(\cos\frac{\xi}{2}e^{-i\frac{\eta}{2}}, \sin\frac{\xi}{2}e^{i\frac{\eta}{2}})^{\rm T}$ and $|\Psi_2\rangle=(\sin\frac{\xi}{2}e^{-i\frac{\eta}{2}}, -\cos\frac{\xi}{2}e^{i\frac{\eta}{2}})^{\rm T}$ following the NNGQC theory, where $(\xi,~\eta)$ determine the vector point on the Bloch sphere. The geometric phase is relevant to the solid angle enclosed by the trajectory and the geodesic. As an example, the gate $U_1=U_{c}(\pi/2,-\pi/2,0)$ is obtained by setting $\phi_0=-\pi/2,\phi_1=\pi/2$, $\xi_0=\pi/2$ and $\tau=\pi/\Omega_0$, with other parameters $\eta_+=-\pi/2$, $\eta_-=\pi/2$, $\xi_+=0$, $\xi_-=\pi$ and the calculated $\gamma=\int_{0}^{\tau}i\langle\Psi_{1}(t)|(d/dt)|\Psi_{1}(t)\rangle=\pi/4$. In contrast, the Hadamard gate is achieved by setting $\phi_0=0$, $\phi_1=\pi/2$, $\xi_0=\pi/2$ and $\tau=3\pi/2\Omega_0$, with other parameters $\eta_+=\pi/2$, $\eta_-=\pi/2$, $\xi_+=\pi/2$, $\xi_-=3\pi/2$ and the calculated $\gamma=\pi/4$. With the given parameters, the trajectory and the geodesic connection for $U_{1}$ gate and Hadamard gate are plotted in Fig.~\ref{figbloch}.

\section{Some details for the gates by nonadiabatic geometric quantum computation}

To achieve a conventional NGQC, the evolution time $\tau$ is divided into three intervals,
\begin{eqnarray}
\int_0^{\tau_1}\Omega(t)dt&=&\theta, \quad \ \ \quad \varphi=-\phi+\frac{\pi}{2}, \qquad t\in [0,\tau_1] \notag \\
\int_{\tau_1}^{\tau_2}\Omega(t)dt&=&\pi, \quad \ \ \quad \varphi=-\phi-\gamma-\frac{\pi}{2}, t\in [\tau_1,\tau_2] \notag \\
\int_{\tau_2}^{\tau}\Omega(t)dt&=&\pi-\theta, \quad \varphi=-\phi+\frac{\pi}{2}, \qquad t\in [\tau_2,\tau] \notag
\end{eqnarray}
with $\theta\in[0,\pi]$. Therefore, the obtained single-qubit gate after the total geometric evolution process is given by
\begin{eqnarray}
U_c(\theta,\gamma,\phi)&=&\exp(i\gamma\vec{n}\cdot\vec{\sigma}) \notag \\
&=&\cos\gamma+i\sin\gamma\cos\theta\sigma_z+i\sin\gamma\sin\theta\cos\phi\sigma_x \notag\\
&&+i\sin\gamma\sin\theta\sin\phi\sigma_y
\end{eqnarray}
which corresponds to a rotation around the axis $\vec{n}=(\sin\theta\cos\phi,\sin\theta\sin\phi,\cos\theta)$ by an angle $-2\gamma$.

In our case, the $U_1$ gate is written as $U_1=[1+i, -1-i;1-i, 1-i]/2=(I+i\sigma_z)/2-i(\sigma_x+\sigma_y)/2$. To construct the $U_1$ gate, we require the parameters to satisfy $U_c(\theta,\gamma,\phi)=U_1$ which means
\begin{equation}
\cos\gamma=\frac{1}{2}, \cos\theta=\frac{1}{\sqrt{3}},\sin\phi=-\frac{1}{\sqrt{2}},\cos\phi=-\frac{1}{\sqrt{2}}. \notag
\end{equation}
Then the parameters are determined as
\begin{equation}
\gamma=\frac{\pi}{3},\theta=\arccos\frac{\sqrt{3}}{3},\phi=\frac{5\pi}{4}.
\end{equation}
Choosing the constant Rabi frequency, i.e., $\Omega(t)=\Omega_m$ with $\Omega_m$ denoting the constant Rabi frequency, we obtain the total time of a quantum gate to be $\tau=2\pi/\Omega_m$.

\section{Some details for the gates by dynamical Quantum Computation}

In our experiment, the DQC is carried out based on the evolution of the qubit state under carrier transitions, as described by the Hamiltonian,
\begin{equation}
H=\frac{1}{2}\Omega_0e^{i\varphi}|e\rangle\langle g|+\rm H.C.,
\end{equation}
with $\Omega_0$ and $\varphi$ denoting the Rabi frequency and the phase of the laser. The corresponding unitary evolution operator is
\begin{equation}
U_d(\theta,\varphi)=\cos\frac{\theta}{2}-i\sin\frac{\theta}{2}(\sigma_+e^{i\varphi}+\sigma_-e^{-i\varphi}),
\end{equation}
with $\theta=\Omega_0 t$. Thus we generate the $U_1$ gate by following pulse sequences
\begin{equation}
U_1=U_d(\frac{\pi}{2},0)U_d(\pi,-\frac{\pi}{4})U_d(\pi,\frac{\pi}{2}),
\end{equation}
which means that the process needs a time duration of $\tau=5\pi/2\Omega_0$.

\section{Tomography of a single qubit state}

Using the Stokes parameters, the arbitrary state of a single qubit can be uniquely represented as
\begin{equation}
\rho=\frac{1}{2}\sum_{i=0}^3 S_i \sigma_i
\end{equation}
which is constituted by Pauli operators, i.e., $\sigma_0=I$, $\sigma_1=\sigma_x$, $\sigma_2=\sigma_y$, and $\sigma_3=\sigma_z $. We have $S_i= \rm Tr[\sigma_i\rho]$ with $ S_0=1$, and three parameters as below need to be measured.
\begin{equation}
S_1=P_{e}^x-P_{g}^x, \quad S_2=P_{e}^y-P_{g}^y, \quad S_3=P_{e}-P_{g},
\end{equation}
where P$^x$ is a projection measurement on $|g\rangle^x=\frac{1}{\sqrt{2}}(|e\rangle-|g\rangle)$ and $|e\rangle^x=\frac{1}{\sqrt{2}}(|e\rangle+|g\rangle)$. P$^y$ is a projection measurement on $|g\rangle^y=\frac{1}{\sqrt{2}}(|e\rangle-i|g\rangle)$ and $|e\rangle^y=\frac{1}{\sqrt{2}}(|e\rangle+i|g\rangle)$. Then the tomography of a qubit state under the carrier transition is carried out by following steps,\\

\noindent\textbf{I. Measurement of $ S_3 $}: Measure $\rho$ in the state $|e\rangle$ to obtain $P_e$; Then swap the states $|g\rangle$ and $|e\rangle$ by applying a resonant $\pi$ pulse and then measure $\rho$  in the state $|g\rangle$ to obtain $P_g$. Finally, calculate $S_3$ using the relation $ S_3= P_e-P_g$. \\

\noindent\textbf{II. Measurement of $ S_1 $}: Firstly, rotate $\rho$ by applying a resonant pulse with $\theta=\phi=\pi/2$ (i.e., $U_x$). Secondly, make the measurement in $|e\rangle$ to obtain $P_e^x$; Then swap the states $|g\rangle $ and $|e\rangle$ by applying a resonant $\pi$ pulse, and then make the measurement in $|e\rangle$ to acquire $P_g^x$. Finally, calculate $S_1$ using the relation $S_1= P_e^x-P_g^x$. \\

\noindent\textbf{III. Measurement of $ S_2 $}: First rotate $\rho$ by applying a resonant pulse with $\theta=\pi/2,\phi=0 $ (i.e., $U_y$). Then make the measurement in $|e\rangle$ and $|g\rangle$ to obtain $P_e^y$ and $P_g^y$, respectively. Then, calculate $S_1$  using the relation $ S_2= P_e^y-P_g^y$. \\

Assuming that the ideal and experimental states are defined by the vector $\vec{S}^{k}\equiv (S_0^k,S_1^k,S_2^k,S_3^k)^T$ with $k=exp, id$, respectively, regarding $\rho_{k}=\frac{1}{2}\sum_{i=0}^3 S^{k}_i \sigma_i$, we have the fidelity of the quantum gate calculated as
\begin{equation}
\mathcal{F}=\text{Tr}[\rho_{\rm exp}\rho_{\rm id}]=\frac{1}{2} \vec{S}^{\rm exp}\cdot \vec{S}^{\rm id}.
\end{equation}
Correspondingly, the infidelity is given by
\begin{equation}
\epsilon_{\mathcal{F}}=\sqrt{\frac{1}{4} \sum_{k=0}^3 (d\vec{S}^{\rm exp}_k \cdot\vec{S}^{\rm id}_k)^2},
\end{equation}
where the vector $d\vec{S}^{\rm exp}$ denotes the experimental error vector of $ \vec{S}^{\rm exp}$.

\section{Tomography for the quantum gate process}

A quantum operation process can be described as  \cite{jmo-44-2455}
\begin{equation}
\varepsilon(\rho)=\sum_{mn}\chi_{mn}A_m\rho A_n^{\dagger},
\end{equation}
where $\chi_{mn}$ is a process matrix and the operators $A_m$ consist of a fixed operator set to describe the process operator $\varepsilon$. In the one-qubit case, the operator set is defined as
\begin{equation}
A_0=I,\ A_1=\sigma_x,\ A_2=-i\sigma_y, \ A_3=\sigma_z.
\end{equation}
Then, the process matrix $\chi$ can be measured by making a tomography of the density matrix for the corresponding output states by choosing four different input states,
\begin{equation}
|g\rangle,\ |e\rangle,\ |+\rangle=(|g\rangle+|e\rangle)/\sqrt{2},\ |-\rangle=(|g\rangle-i|e\rangle)/\sqrt{2}.
\end{equation}
The four density matrices are determined by the following forms
\begin{eqnarray}
\rho_1&=&\varepsilon(|g\rangle\langle g|), \notag \\
\rho_2&=&\varepsilon(|e\rangle\langle e|), \notag \\
\rho_3&=&\varepsilon(|+\rangle\langle +|)-i\varepsilon(|-\rangle\langle -|)-\frac{(1-i)(\rho_1+\rho_2)}{2}, \notag \\
\rho_4&=&\varepsilon(|+\rangle\langle +|)+i\varepsilon(|-\rangle\langle -|)-\frac{(1+i)(\rho_1+\rho_2)}{2}. \notag
\end{eqnarray}
Actually, the states $\rho_{3,4}$ correspond to the output matrices of the input states $|g\rangle\langle e|$ and $|e\rangle\langle g|$.

Therefore, the process matrix $\chi$ can be expressed as
\begin{equation}
\chi=\Lambda
\left[
\begin{matrix}
\rho_1 & \rho_4 \\
\rho_3 & \rho_2
\end{matrix}
\right] \Lambda, \rm{with} \
\Lambda=\frac{1}{2}
\left[
\begin{matrix}
I & \sigma_x \\
\sigma_x & -I
\end{matrix}
\right].
\end{equation}
Then, the fidelity of the gate process for the ideal and experimental results is written as
\begin{equation}
\mathcal{F}_{gp}=\text{Tr}[\chi_{\rm exp}\chi_{\rm id}].
\end{equation}

%\begin{figure}[tbph]
%\centering
%\includegraphics[width=8.0cm,height=2.0 cm]{Fig48.png}
%\includegraphics[width=8.0cm,height=4.0 cm]{Ramsey.eps}
%\caption{(Upper) The Ramsey sequence to measure the dephasing rate in the experiment where the initial sate of system is prepared in the ground state $|g\rangle$. (Bottom) The numerical and analytical results of population in the excited state $|e\rangle$ after the Ramsey evolution where the parameters are selected as $\Gamma=500$ $\mu$s, $\Omega/2\pi=100$ kHz, $\Delta=3\Omega/200$ and $p=0.96$. }
%\label{Fig2s3}
%\end{figure}

\section{Supplemental figures regarding the main text}

We present in Figs. 3-6 the experimental data of the Stokes parameters for different output states in the gate processes by different schemes.

\begin{figure}[tbph]
	\centering
	\includegraphics[width=8.5cm,height=3.5 cm]{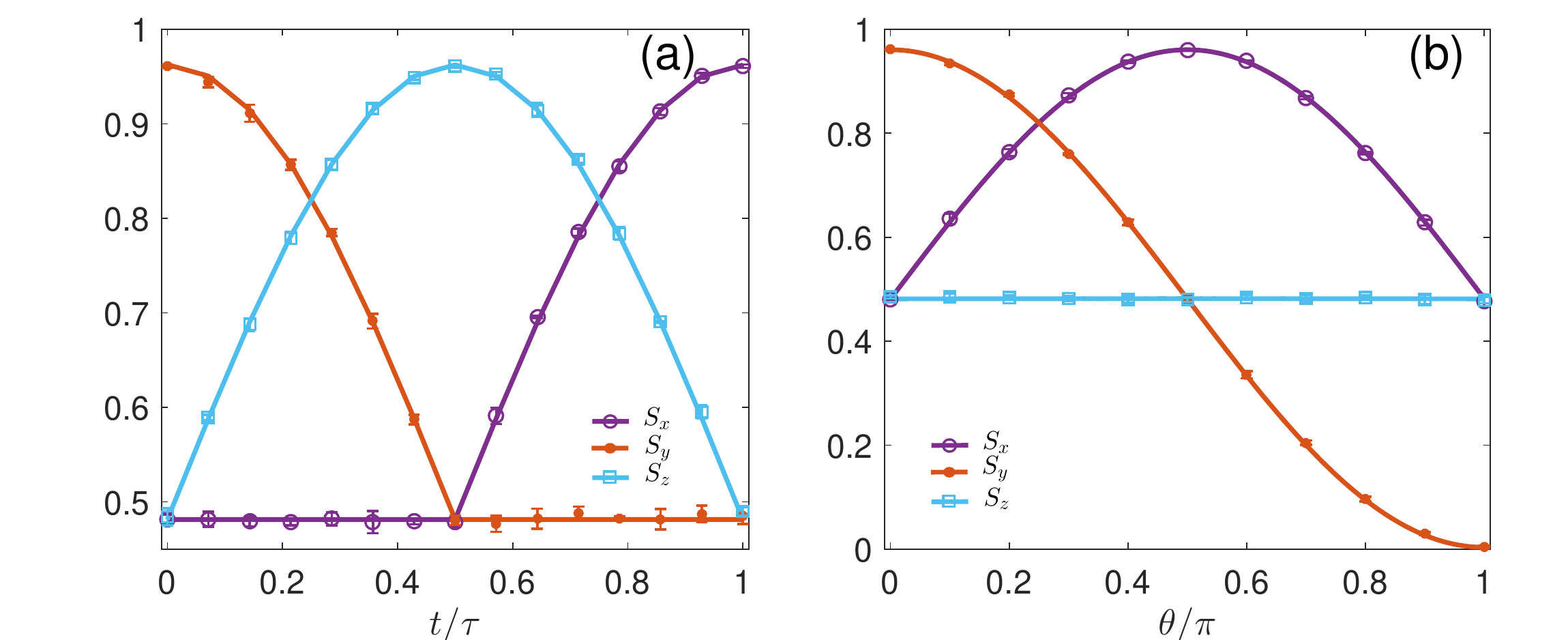}
	\caption{Stokes parameters $S_k$ ($k=x,y,z$) regrading the experimental data of Fig. 1 in the main text. (a) For the evolution process of the gate under different time durations (main text Fig. 1(c)). (b) For the output states with different input states  (main text Fig. 1(d)).}
	\label{Figs1}
\end{figure}

\begin{figure}[tbph]
	\centering
	\includegraphics[width=8.5cm,height=3.5 cm]{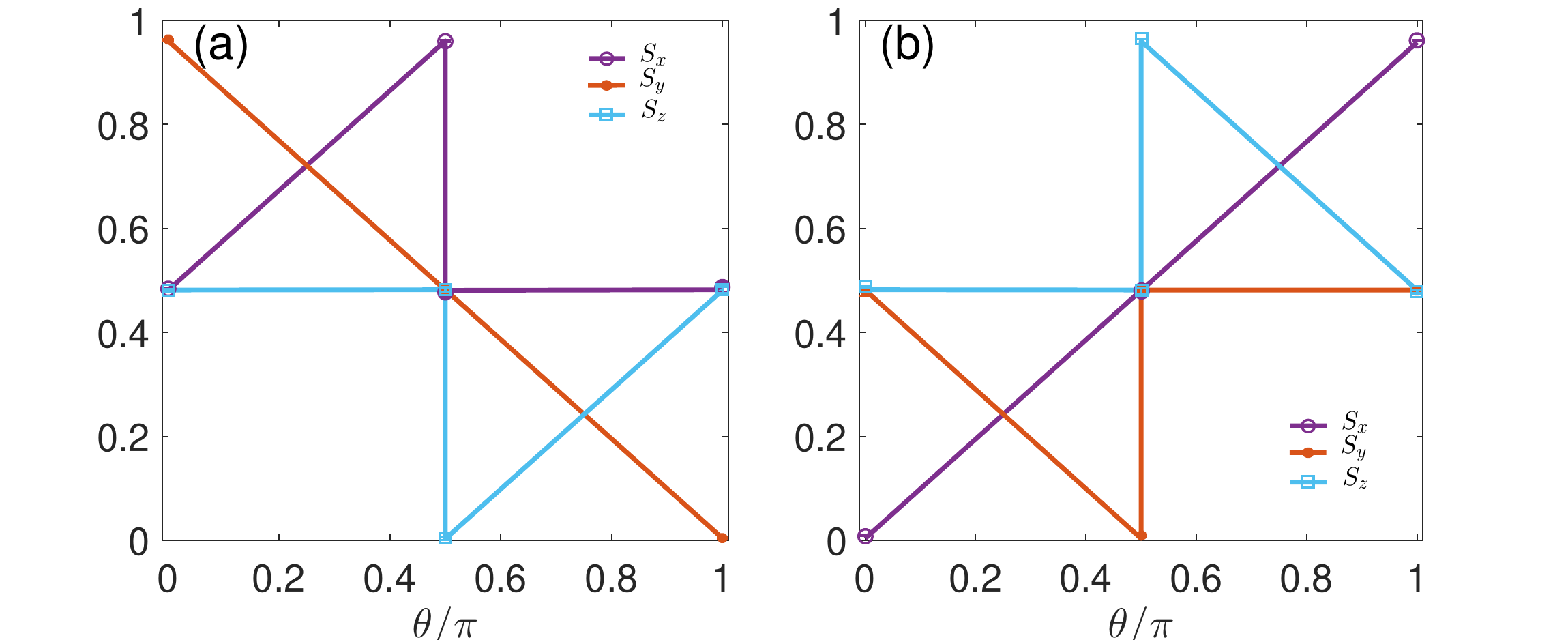}
	\caption{Stokes parameters $S_k$ ($k=x,y,z$) regrading the experimental data of Fig. 2 in the main text. (a) For the $U_1$ gate process and (b) For the Hadamard gate process under different input states. }
	\label{Figs2}
\end{figure}

\begin{figure*}[tbph]
	\centering
	\includegraphics[width=16.0cm,height=4.0 cm]{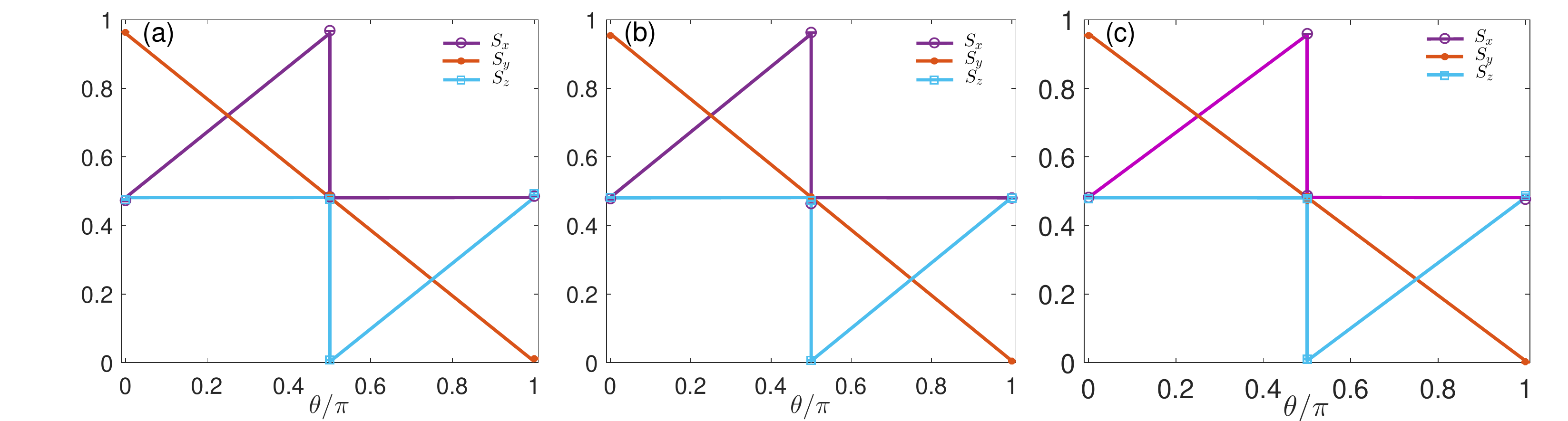}
	\caption{Stokes parameters $S_k$ ($k=x,y,z$) for the experimental data of $U_1$ gate process in Fig. 3 of the main text. (a-c) For NNGQC, NGQC and DQC, respectively.  }
	\label{Figs3}
\end{figure*}

\begin{figure*}[tbph]
	\centering
	\includegraphics[width=16.0cm,height=7.0 cm]{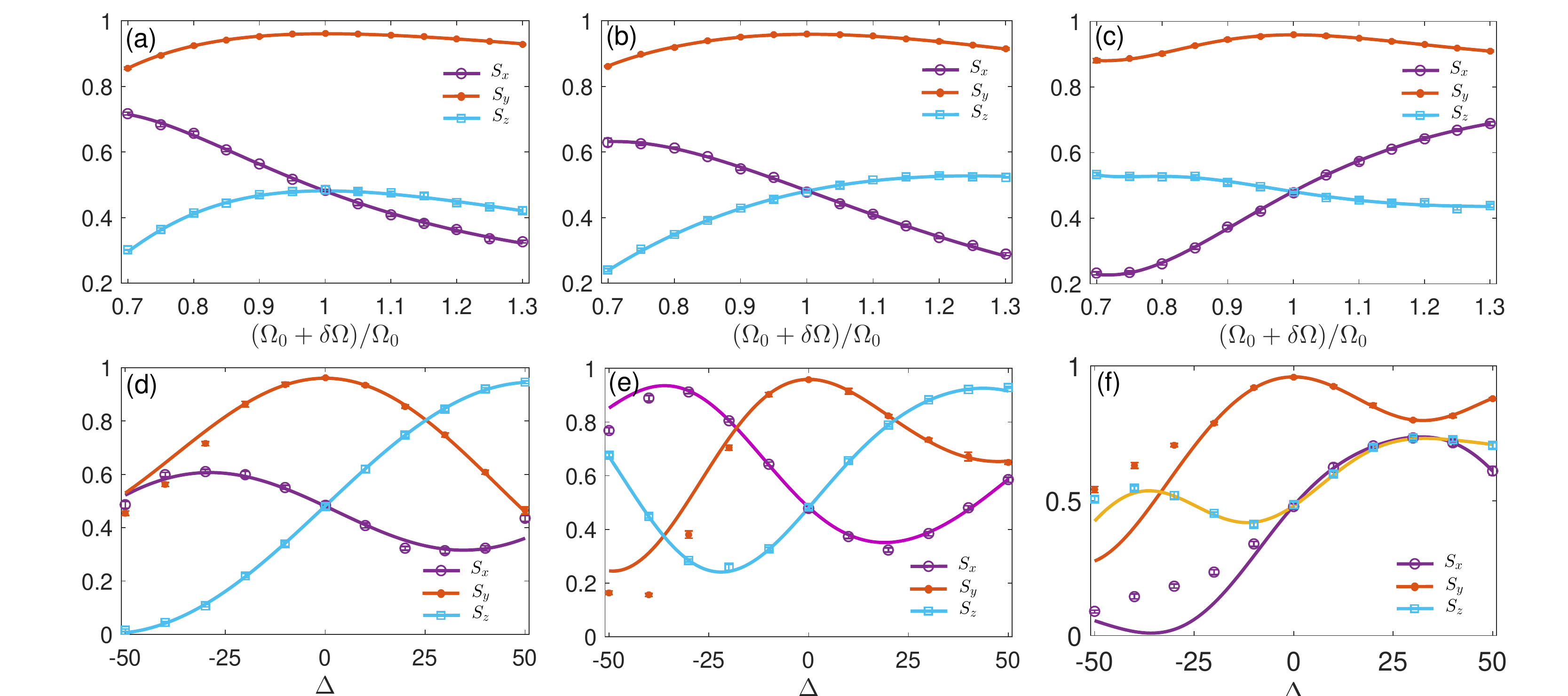}
	\caption{Stokes parameters $S_k$ ($k=x,y,z$) for the experimental data of $U_1$ gate process in Fig. 4 of the main text. (a-c) Regarding different Rabi frequency errors for NNGQC, NGQC and DQC, respectively, and (d-f) Regarding different resonance frequency errors corresponding to NNGQC, NGQC and DQC, respectively. }
	\label{Figs4}
\end{figure*}

\section{Scheme for the two-qubit NNGQC} \label{T}

We elucidate below that the two-qubit NNGQC could be essentially carried out by the similar way to the single-qubit counterpart.

\begin{figure}[tbph]
	\centering
	\includegraphics[width=8.0cm,height=3.0 cm]{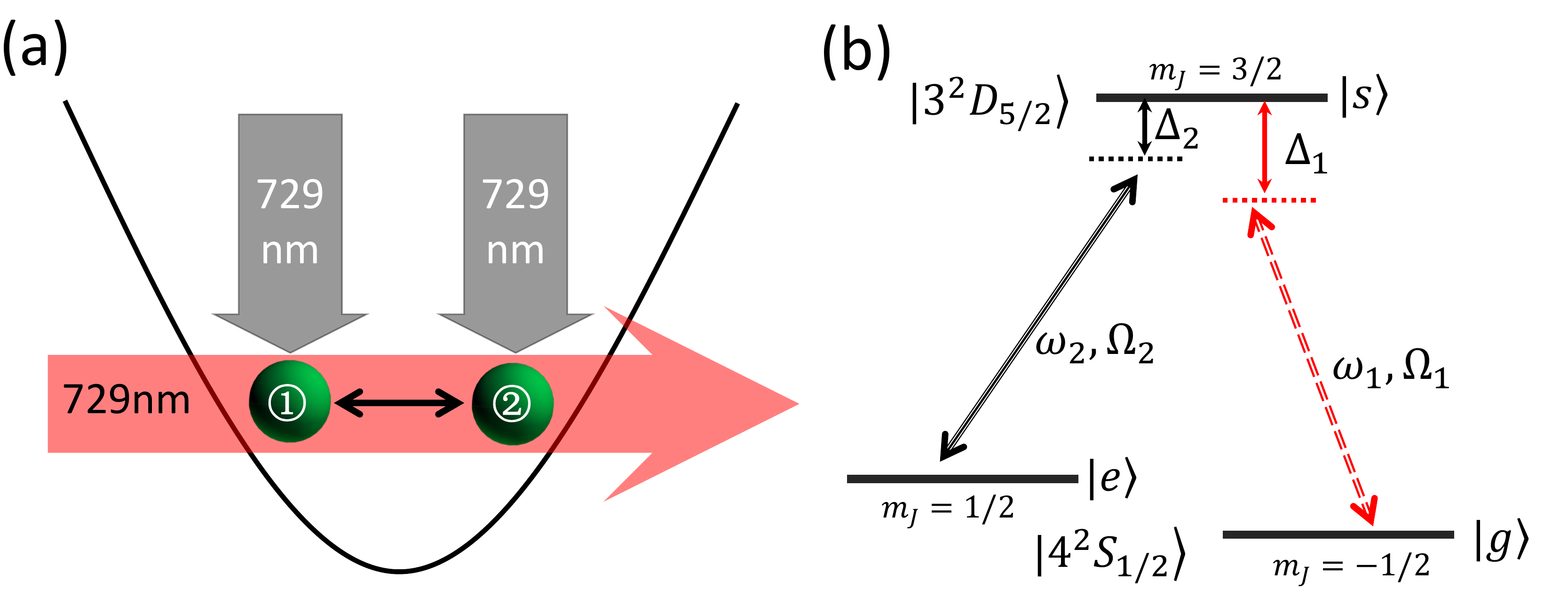}
	\caption{(a) Schematic of two trapped ions driven by two 729-nm lasers, where one laser (red beam) irradiates along the ions' vibrational direction, i.e., $\theta_1=0$ and the other irradiation (black beams) perpendicular to the vibrational direction (i.e., $\theta_2=\pi/2$) with the requirement of individually addressing. (b) Level scheme of the ion, where the Zeeman sublevels of S$_{1/2}$ are labeled as $|g\rangle$ and $|e\rangle$, and one Zeeman level $m_{J}=3/2$ of D$_{5/2}$ is employed as the auxiliary state $|s\rangle$.}
	\label{Figsp}
\end{figure}

Consider two ions simultaneously driven by two 729-nm lasers and coupled by the vibrational mode of motion. As plotted in Fig. \ref{Figsp}, the two lasers couple the levels $|g\rangle$ and $|e\rangle$ to the auxiliary level $|s\rangle$ with the corresponding detunings $\Delta_{1,2}$ and Rabi frequencies $\Omega_{1,2}$, which can be described, in the interaction picture, by the Hamiltonian as
\begin{equation}
H_I=\sum_{j=1}^2\Omega_{j}\sigma_+^je^{-i\Delta_j t+i\phi_j}\exp[i\tilde{\eta}_j(a^{\dagger}e^{i\omega t}+ae^{-i\omega t})]+H.C. ,
\label{EqT1}
\end{equation}
where $\omega$ is the vibrational frequency of center-of-mass mode, $\phi_j$ denotes the laser phase, $\sigma_+^1=|s\rangle\langle g|$, $\sigma_+^2=|s\rangle\langle e|$, $\Delta_1=\omega_1+\omega_g-\omega_s$, $\Delta_2=\omega_2+\omega_e-\omega_s$ and the Lamb-Dicke parameters are defined as $\tilde{\eta}_j=k\cos\theta_j\sqrt{\hbar/2m\omega}$ with $\theta_j$ denoting the intersection angle between the $j$-th laser irradiation and the vibrational direction. In our case, we choose $\theta_1=0$ and $\theta_2=\pi/2$. By setting $\Delta_1=-\omega-\Delta$ and $\Delta_2=-\Delta$ with $\Delta\ll\omega$, we rewrite the above Hamiltonian, under the rotating wave approximation ($\Omega_{1,2}\ll\omega$), as
\begin{equation}
H_I=(\Delta |s\rangle+\Omega e^{-i\bar{\phi}}|e\rangle+\frac{g}{2}a^{\dagger}|g\rangle)\langle s|+H.C. ,
\label{EqT2}
\end{equation}
where $\Omega=\Omega_{2}$, $g=2\eta_1\Omega_{1}$ and we have set $\phi_1=-\pi/2, \phi_2=\bar{\phi}$. Meanwhile, the scheme requires that the phase of the second laser on different ions can be independently adjusted. Thus, the total Hamiltonian of the system can be written as
\begin{equation}
H_t=\sum_{k=1}^2(\Delta |s\rangle_k+\bar{\Omega}_k|e\rangle_k+\frac{g}{2}a^{\dagger}|g\rangle_k)\langle s|+H.C.
\label{EqT3}
\end{equation}
where $\bar{\Omega}_k=\Omega e^{-i\bar{\phi}_k}$ and $k$ denotes the $k$th ion.

Given the system in the ground state of vibration, the initial state of the system will be restricted in the space spanned by {$|\psi_1\rangle=|ee0\rangle, |\psi_2\rangle=|eg0\rangle, |\psi_3\rangle=|ge0\rangle,|\psi_4\rangle= |gg0\rangle$}, where the first and the second terms of the ket denotation denote the state of two ions and the third term represents the phonon states of the system. When the system is initially in $|\psi_1\rangle$, the evolution will occur in the subspace $C_{ee0}=\{|\psi_1\rangle,|\psi_5\rangle,|\psi_6\rangle,...,|\psi_{12}\rangle\}$ with
\begin{eqnarray}
|\psi_5\rangle&=&|se0\rangle,\ |\psi_6\rangle=|es0\rangle,\
|\psi_7\rangle=|ge1\rangle,\ |\psi_8\rangle=|eg1\rangle\notag\\
|\psi_9\rangle&=&|ss0\rangle,\ |\psi_{10}\rangle=|gs1\rangle,
\ |\psi_{11}\rangle=|sg1\rangle,\ |\psi_{12}\rangle=|gg2\rangle.\notag
\end{eqnarray}
The Hamiltonian of $C_{ee0}$ can be written as $\hat{\mathcal{H}}_{ee0}=\hat{\mathcal{H}}_{g}+\hat{\mathcal{H}}_{ \bar{\Omega}}+\hat{\mathcal{H}}_{\Delta}$ with
\begin{eqnarray}
\hat{\mathcal{H}_{g}}&=& g(\left| \psi_{5}\right\rangle  \left\langle \psi_{7}\right|+\left| \psi_{6}\right\rangle  \left\langle \psi_{8}\right| +\left| \psi_{9}\right\rangle  \left\langle \psi_{10}\right|+\left| \psi_{9}\right\rangle  \left\langle \psi_{11}\right|\notag\\
& &+\left| \psi_{10}\right\rangle  \left\langle \psi_{12}\right|+\left| \psi_{11}\right\rangle  \left\langle \psi_{12}\right|)+\rm H.c. \notag \\
\hat{\mathcal{H}}_{\bar{\Omega}}&=&\bar{\Omega}_{1}(\left| \psi_{1}\right\rangle  \left\langle \psi_{5}\right|+\left| \psi_{1}\right\rangle  \left\langle \psi_{9}\right|+\left| \psi_{8}\right\rangle  \left\langle \psi_{11}\right|)\notag\\
& &+\bar{\Omega}_{2}(\left| \psi_{1}\right\rangle  \left\langle \psi_{6}\right|
+\left| \psi_{5}\right\rangle  \left\langle \psi_{9}\right|+\left| \psi_{7}\right\rangle  \left\langle \psi_{10}\right| )+{\rm H.c.}\notag\\
\hat{\mathcal{H}}_{\Delta}&=&\Delta(\left| \psi_{5}\right\rangle  \left\langle \psi_{5}\right|+\left| \psi_{6}\right\rangle  \left\langle \psi_{6}\right|+\sum_{j=9}^{11}\left| \psi_{j}\right\rangle  \left\langle \psi_{j}\right|).
\end{eqnarray}
For a quantum Zeno dynamical process under the condition $\hat{\mathcal{H}}_{g}\gg\ \hat{\mathcal{H}}_{\bar{\Omega}}$, i.e., $g\gg\bar{\Omega}_{1,2}$, the system will evolve within the subspace of the initial eigenstates, and the bosonic modes have three zero-eigenvalue eigenstates \{ $\left| \Phi_{d1}\right\rangle=(\left| \psi_{9}\right\rangle+\left| \psi_{12}\right\rangle)/\sqrt{2} $, $\left| \Phi_{d2}\right\rangle=(\left| \psi_{10}\right\rangle+\left| \psi_{11}\right\rangle  )/\sqrt{2} $, $\left| \psi_{1}\right\rangle $\}. Therefore, there is no effective coupling among $\left| \Phi_{d1}\right\rangle$, $\left| \Phi_{d2}\right\rangle$ and $\left| \psi_{1}\right\rangle $ and the system will be trapped in the initial state $|\psi_1\rangle=|ee0\rangle$ all the time.

If the initial state is $|\psi_{2}\rangle$ or $|\psi_{3}\rangle$, the evolution will be remaining in the subspace $C_{eg0}=\{|\psi_2\rangle, |\psi_3\rangle,|\psi_{13}\rangle,|\psi_{14}\rangle,|\psi_{15}\rangle\}$ with
\begin{equation}
|\psi_{13}\rangle=|eg0\rangle,\ |\psi_{14}\rangle=|sg0\rangle,\  |\psi_{15}\rangle=|gg1\rangle, \notag
\end{equation}
the Hamiltonian in the subspace $C_{eg0}$ can be expressed as $\hat{H}_{eg0}=\hat{H}_{g}+\hat{H}_{\bar{\Omega}}+\hat{H}_{\Delta} $ with
\begin{eqnarray}
\hat{H}_{g}&=& g(\left| \psi_{13}\right\rangle  \left\langle \psi_{14}\right|+\left| \psi_{14}\right\rangle  \left\langle \psi_{15}\right| +\rm H.c., \notag\\
\hat{H}_{\bar{\Omega}}&=&\bar{\Omega}_{1}\left| \psi_{2}\right\rangle  \left\langle \psi_{13}\right|+\bar{\Omega}^{*}_{2}\left| \psi_{15}\right\rangle  \left\langle \psi_{3}\right|+{\rm H.c.},\\
\hat{H}_{\Delta}&=&\Delta(\left| \psi_{13}\right\rangle  \left\langle \psi_{13}\right|+\left| \psi_{15}\right\rangle  \left\langle \psi_{15}\right|).\notag
\end{eqnarray}

Given the initial state $\left| \psi_{4}\right\rangle=\left| gg0\right\rangle $, the system keeps unchanged based on the Hamiltonian in Eq.~(\ref{EqT3}).

Thus, in the picture of the eigenstates of $\hat{H}_g$: $\{\left| \Psi_{\pm}\right\rangle=(\left| \psi_{13}\right\rangle\pm \sqrt{2}\left| \psi_{14}\right\rangle +\left| \psi_{15}\right\rangle)/2,\ \left| \Psi_{d}\right\rangle=(\left| \psi_{13}\right\rangle -\left| \psi_{15}\right\rangle)/\sqrt{2}\}$ corresponding to the eigenvalues $\pm\sqrt{2}g$ and 0, the Hamiltonian can be represented as
\begin{eqnarray}
\hat{H}_{\bar{\Omega}}&=&\bar{\Omega}_{1}\left| \psi_{2}\right\rangle\left\langle \Phi_{+}\right| +\bar{\Omega}_{2}\left|\psi_{3}\right\rangle \left\langle \Phi_{-}\right|+\rm H.c., \notag\\
\hat{H}_{\Delta}&=&2\Delta\left| \Psi_{d}\right\rangle\left\langle \Psi_{d}\right|+\Delta\left| \Phi_{+}\right\rangle   \left\langle \Phi_{-}\right|+\rm H.c.,
\end{eqnarray}
where the shorthand states $\left| \Phi_{\pm}\right\rangle\equiv(e^{i\sqrt{2}gt}\left| \Psi_{+}\right\rangle +e^{-i\sqrt{2}gt}\left| \Psi_{-}\right\rangle \pm \sqrt{2}\left| \Psi_{d}\right\rangle)/2$. Due to the condition of $g\gg\bar{\Omega}_{1,2}$, the states $ \left| \Psi_{\pm} \right\rangle $ will be decoupled from \{$\left| \Psi_{d} \right\rangle, \left| \psi_{2} \right\rangle, \left| \psi_{3} \right\rangle$\},  and then the effective Hamiltonian is obtained as
\begin{equation}
\hat{H}_{\rm eff}=\Delta \left| \Psi_{d}\right\rangle\left\langle \Psi_{d}\right|+\frac{1}{\sqrt{2}}(\bar{\Omega}_{1}\left| \psi_{2}\right\rangle-\bar{\Omega}_{2}\left|\psi_{3}\right\rangle )\left\langle \Psi_{d}\right|+\rm H.c. \notag
\end{equation}
Under the large detuning condition $\Delta\gg \bar{\Omega}_{1,2}$ and using the James-Jerke method \cite{JJ}, the above equation can be rewritten as
\begin{eqnarray}
\hat{H}_{\rm eff}&=&\frac{|\bar{\Omega}_{1}|^2}{2\Delta}(\left| \psi_{2}\right\rangle  \left\langle \psi_{2}\right|-\left| \Psi_{d}\right\rangle  \left\langle \Psi_{d}\right|)+\frac{|\bar{\Omega}_{2}|^2}{2\Delta}(\left| \psi_{3}\right\rangle  \left\langle \psi_{3}\right|\notag\\
& &-\left| \Psi_{d}\right\rangle  \left\langle \Psi_{d}\right|)-\frac{\bar{\Omega}_{1}^*\bar{\Omega}_{2}}{2\Delta}\left| \psi_{3}\right \rangle  \left\langle \psi_{2}\right|+\rm H.c.,
\end{eqnarray}
where the first two terms are caused by the Stark shifts and $\left| \Psi_{d}\right \rangle$ is decoupled from $\left| \psi_{2}\right \rangle$ and $\left| \psi_{3}\right \rangle$. Choosing $\widetilde{\Omega}\equiv\lvert \bar{\Omega}_{1}\rvert=\lvert \bar{\Omega}_{2}\rvert$ and
ignoring the global phase factor, the effective Hamiltonian in the interaction picture can be further reduced to
\begin{equation}\label{effhamiltonian}
\hat{H}_{\rm eff}=\frac{\widetilde{\Omega}^{2}e^{i\varphi}}{2\Delta}\left| \psi_{3}\right\rangle  \left\langle \psi_{2}\right|+\rm H.c.
\end{equation}
with the time-dependent phase difference $\varphi=\pi+\bar{\phi}_{1}-\bar{\phi}_{2}$. This Hamiltonian is actually of the similar form to Eq. (1) in the main text, which in principle can be carried out by the steps of NNGQC, and enjoy the advantages as we verified. Considering $\hat{U}'(\theta,~\alpha,~\beta)$ as the evolution operator relevant to Eq. (24), we may have
the two-qubit evolution operator, which is unitarily equivalent to $\hat{U}'(\theta,~\alpha,~\beta)$, constituting a universal two-qubit controlled gate within the subspace spanned by
$|\psi_1\rangle$, $|\psi_2\rangle$, $|\psi_3\rangle$ and $|\psi_4\rangle$. Note that $\hat{U}'(\theta,~\alpha,~\beta)$ owns the same matrix form as $\hat{U}(\theta,~\alpha,~\beta)$ in the main text, but only works on the states $|\psi_2\rangle$ and $|\psi_3\rangle$. $|\psi_1\rangle$ and $|\psi_4\rangle$ remain unchanged in this two-qubit controlled gate. The NNGQC operational details as well as the advantages regarding higher-speed, higher-fidelity and lower error in comparison with the same gating operations by other ways under the same conditions/paramters will be published elsewhere.


\begin{thebibliography}{99}
\bibitem{explain} The two-qubit NNQGC gate designed in \cite{SM} is built on the qubit encoded in the Zeeman sublevels of the S$_{1/2}$, while our single-qubit NNQGC gate discussed above is executed between S$_{1/2}$ and $D_{5/2}$. However, the single-qubit NNQGC gate can also be performed between the Zeeman sublevels of the S$_{1/2}$, whose operations are mathematically equivalent to what were described in the present work. Therefore, we consider that our experimental witness, as a proof-of-principle work, is a full verification of the advantages of both the single-qubit and two-qubit NNQGC gates.
\end{thebibliography}
\end{document}